\documentclass[a4paper,11pt]{article}
\pdfoutput=1 

\usepackage{jcappub} 

\usepackage[T1]{fontenc} 
\usepackage{amsmath}
\usepackage{amsmath, bm}
\usepackage{amssymb}
\allowdisplaybreaks

\usepackage{makeidx}
\usepackage{amsfonts}
\usepackage[usenames,dvipsnames]{pstricks}
\usepackage{subfigure}
\usepackage{epsfig}
\usepackage{pst-grad} 
\usepackage{mathtools}
\usepackage{array}
\usepackage{tikz}
\usepackage{flushend}
\newcommand\nn{{\nonumber}}
\newcommand{\cdd}{\mathrm{D}}

\title{Effect of Particle Spin on Trajectory Deflection and Gravitational Lensing}

\author[a,1]{Zhuoming Zhang,\note{These authors contributed equally to this work.}}
\author[a,1]{Gaofeng Fan}
\author[b,2]{and Junji Jia\note{Corresponding author.}}

\affiliation[a]{School of Physics and Technology, Wuhan University,\\Wuhan, 430072, China}
\affiliation[b]{Center for Astrophysics \& MOE Key Laboratory of Artificial Micro- and Nano-structures,\\School of Physics and Technology, Wuhan University, Wuhan, 430072, China}

\emailAdd{zhuoming\_zhang@whu.edu.cn}
\emailAdd{gaofeng\_fan@whu.edu.cn}
\emailAdd{junjijia@whu.edu.cn}

\abstract{Spin of a test particle is a fundamental property that can affect its motion in a gravitational field. In this work we consider the effect of particle spin on its deflection angle and gravitational lensing in the equatorial plane of arbitrary stationary and axisymmetric spacetimes. To do this we developed a perturbative method that can be applied to spinning signals with arbitrary asymptotic velocity and takes into account the finite distance effect of the source and the observer. The deflection angle $\Delta\varphi$ and total travel time $\Delta t$ are expressed as (quasi-)power series whose coefficients are polynomials of the asymptotic expansion coefficients of the metric functions. It is found that when the spin and orbital angular momenta are parallel (or antiparallel), the deflection angle is decreased (or increased). Apparent angles $\theta$ of the images in gravitational lensing and their time delays are also solved. In Kerr spacetime, spin affects the apparent angle $\theta_K$ in a way similar to its effect on $\Delta\varphi_K$. The time delay between signals with opposite spins is found to be proportional to the signal spin at leading order. These time delays might be used to constrain the spin to mass ratio of neutrinos.}


\arxivnumber{2207.09194}

\begin{document}
\maketitle
\flushbottom

\section{Introduction}

In the study of classical effects of General Relativity (GR), deviation of trajectories of test particles from their Newtonian prediction plays a crucial role. Among these, the deflection of light rays is one of the most important, for it helped the acceptance of GR by both the physics community and general public \cite{Dyson:1920cwa}.
Deflection of light rays is also the basis of gravitational lensing (GL), which has developed into an important tool in astrophysics. GL can be used not only to measure the Hubble constant \cite{Refsdal:1964nw,Kundic:1996tr} and map the mass of superclusters \cite{Hoekstra:2000ux,Gray:2001zx}, but also to explore dark energy and test new gravitational theories \cite{Hoekstra:2008db,Joyce:2016vqv}.

In all these deflection or lensings, the messengers usually are electromagnetic waves of various wavelength. The underlying theory in these phenomena basically always assumed that the messengers (i.e. light rays) are massless and spinless test particles. However, with the discovery of neutrinos from extragalactic sources \cite{Kamiokande-II:1987idp, Bionta:1987qt,IceCube:2018dnn,IceCube:2018cha} and the GL of supernovae \cite{Kelly:2014mwa,Goobar:2016uuf}, as well as the long known cosmic rays \cite{Letessier-Selvon:2011sak}, it is clear that timelike signals with spin can also experience the gravitational deflection and act as messengers in GL. To reveal information about the messengers and the spacetimes from GL of these signals, therefore one has to study the effect of spin as well as the timelike nature of these signals on their deflection.

Theoretically, in recent years different methods have been developed to compute the deflection of massive signals. These methods, including the Gauss-Bonnet theorem method \cite{Gibbons:2008rj,Crisnejo:2018uyn,Li:2019qyb} and perturbative method \cite{Jia:2020xbc,Huang:2020trl}, might be used to spacetimes with different symmetries or generalities \cite{Crisnejo:2019ril,Liu:2020wcu}, or to include the finite distance effect of the source and detector \cite{Ishihara:2016vdc,Li:2019qyb,Huang:2020trl}, and to handle the extra Lorentz force from electromagnetic field \cite{Crisnejo:2018uyn,Crisnejo:2019ril,Li:2020ozr,Xu:2021rld,Zhou:2022dze}.

Now regarding the effect of spin in gravity, there are enormous amount of works on spin's effect in binary systems of compact objects with comparable masses \cite{Kidder:1992fr,Kidder:1995zr, Campanelli:2007ew}, especially on their gravitational wave signature \cite{Lang:2006bsg,Arun:2008kb,Farr:2017uvj}. If we focus on the motion of test particles in gravity of a much heavier body, although there are plenty of works on the formalism about how to describe spinning particles in gravity \cite{Mathisson:1937zz,Papapetrou:1951pa,Dixon1964,Hanson:1974qy,Obukhov:2013zca,Cotaescu:2014jca,Lukes-Gerakopoulos:2014dma,Chavanis:2017loo,Costa:2017kdr},
effect of spin on (inner-most stable) circular orbits  \cite{Plyatsko:2010zz,Zhang:2017nhl,Zhang:2018eau,Shaymatov:2021nff}, precession of spin \cite{Pomeransky:2000pb,Iorio:2011ubn,Chakraborty:2016mhx,Ruangsri:2015cvg,Rizwan:2018rgs} and some general features about spinning particles' trajectories (stability \cite{Mohseni:2010rm}, geodesicity \cite{Hojman:2018evi} etc.), the effect of signal spin on deflection and GL has been seldom studied so far. Vines \cite{Vines:2017hyw}, Bini and Geralico \cite{Bini:2018zxp} studied the deflection of spinning signals in Kerr spacetime for ultra-relativistic particles only recently. Barrab\`{e}s and Hogan's work \cite{Barrabes:2003ey} also contains a deflection angle in this spacetime but only to the lowest order.
However, none of these has considered the deflection of spinning signals with subluminal velocity, nor they studied any GL effect.

In this work, we will develop a perturbative method to calculate the deflection angle of spinning signals not only with arbitrary asymptotic velocity but also in arbitrary stationary and axisymmetric spacetimes, and with finite distance effect taken into account. We will also adapt a natural GL equation to systematically study effects of signal spin on their GL, including influences on the apparent angles and time delays of the images. The method and corresponding results are then applied to Kerr and Teo wormhole spacetimes to illustrate their correctness and physical implications.

The work is organized as follows. In section \ref{sec:sasdeflection}, the preliminaries including the basic equations of motion and their transformation in arbitrary stationary and axisymmetric spacetimes are outlined.  Section \ref{sec:dphianddt} is devoted to the perturbative method to compute the deflection angle $\Delta\varphi$ and total travel time $\Delta t$. In section \ref{sec:glgeneral}, the formulas for $\Delta\varphi$ and $\Delta t$ are used in GL of spinning signals to find the effect of spin on the apparent angles and time delays of the images. Section \ref{sec:apps} applies these general results to the Kerr and Teo wormhole spacetimes. We conclude the paper with a discussion in section \ref{sec:conc}.
Throughout the paper, the natural units $G=c=1$ are used.

\section{Equations of motion and their transformation \label{sec:sasdeflection}}
In a spacetime with metric $g_{\mu\nu}$ and Riemann tensor $R^{\mu}_{\nu\rho\sigma}$, the equations of motion of a spinning test particle are the Mathisson–Papapetrou–Dixon equations \cite{Hojman1975}
\begin{align}
\frac{\cdd {p}^{\mu}}{\dd \tau}=&-\frac{1}{2}R^{\mu}_{\nu\rho\sigma}u^{\nu}S^{\rho\sigma}, \label{eq:dotpmu}\\
\frac{\cdd {S}^{\mu\nu}}{\dd \tau}=&2p^{\lsb\mu\right.}u^{\left.\nu\rsb}, \label{eq:dotsmunu}
\end{align}
where $p^{\mu}$, $u^{\mu}\equiv \dd x^\mu/\dd \tau$ and $S^{\mu\nu}$ are respectively the generalized total four-momentum, the four-velocity and the skew spin tensor of the test particle along its trajectory characterized by the proper time $\tau$. The $\cdd/\dd \tau$ denotes the absolute derivative with respect to $\tau$, i.e. for any vector field $T^\mu$
\be
\frac{\cdd {T}^\mu}{\dd \tau}\equiv\frac{\dd T^{\mu}}{\dd {\tau}}+\Gamma^\mu_{\rho\sigma} T^\rho u^\sigma,\label{absolutederivative}\ee
where $\dd/\dd\tau$ stands for the total derivative. Note there are also other equation systems that deal with spinning test particles in gravity \cite{Deriglazov:2017jub}.

The above equations contain fewer constraints than freedoms and therefore to solve them one has to supplement  additional conditions. We will choose the commonly used Tulczyjew-Dixon constraint \cite{Tulczyjew}
\be
S^{\mu\nu}p_{\nu}=0.
\label{eq:smunupnu}
\ee
Using this, and noting $S^{\mu\nu}$ is antisymmetric, we can show that the $S^{\mu\nu}$ and $p^\mu$ satisfy the corresponding normalization conditions  \cite{Zalaquett:2013ifd}
\begin{align}
p^{\mu} p_{\mu}=&-m^{2}\quad \lb m\geq 0\rb, \label{eq:pmupmu}\\
\frac{1}{2} S^{\mu \nu} S_{\mu \nu}=&J_m^{2} \quad \lb J_m\geq 0\rb,\label{eq:12smunusmunu}
\end{align}
where $m$ is the rest mass of the test particle and $J_m$ is the size of its spin angular momentum.
Moreover, if the spacetime contains symmetries described by $n$ Killing vectors $\xi_{\mu}^{\lb k\rb}~\lb k=1,2,\cdots,n\rb$, then there can exist $n$  conserved quantities along the motion
\be
C^{\lb k\rb}=p^{\mu}\xi_{\mu}^{\lb k\rb}-\frac{1}{2}S^{\mu\nu}\xi_{\mu;\nu}^{\lb k\rb} ~\lb k=1,2,\cdots,n\rb. \label{eq:ck}
\ee

Now we consider the realization of the above equations in a stationary and axisymmetric spacetime. Such a spacetime can always be described by the following line element \cite{Sloane:1978ne,Ono:2017pie}
\be
\dd s^2=-A \dd t^2+B\dd t\dd \varphi+D\dd r^2+C\dd\varphi^2 +F \dd \theta^2,
\label{eq:SASlinement}
\ee
where $x^\mu=\lb t,~r,~\theta,~\varphi\rb$ are the coordinates, and $A$, $B$, $C$, $D$, $F$ are metric functions of $r$ and $\theta$ only.
For simplicity, we will only consider test particles moving in the equatorial plane of this spacetime ($\theta(\tau)=\pi/2$ and consequently $u^\theta(\tau)=0$). We will further concentrate on particles whose vectorial spin
\be
S_\mu= \frac{\sqrt{-g}}{2m} \varepsilon_{\mu\alpha\beta\gamma}S^{\alpha\beta} p^\gamma \label{vectorialspin}
\ee
where $g= \det \lb g_{\mu\nu}\rb$ and $\varepsilon_{\mu\nu\alpha\beta}$ is the Levi-Civita symbol,
is perpendicular to the equatorial plane, i.e., $S^t=S^r=S^\varphi=0,~S^\theta\neq0$. Using the constraint \eqref{eq:smunupnu}, this implies that $p^\theta=0$. To satisfy these conditions, we directly choose the solution
\be p^\theta=0=S^{\mu\theta}. \label{eq:zerocomps}
\ee
Indeed, we are able to show that if this condition is true at some $\tau=\tau_0$, then for metrics satisfying the following condition on the equatorial plane
\begin{align}
\partial_\theta A =\partial_\theta B =\partial_\theta C =\partial_\theta D =0,
\end{align}
the $p^\theta(\tau)$ and $S^{\mu\theta}(\tau)$ will remain zero after $\tau_0$. This kind of spacetimes includes many familiar ones such as the Kerr spacetime and Teo wormhole spacetime, which will be studied in section \ref{sec:apps}.

Our goal in the reminder of this section is then to derive the  equations for $\dd\varphi/\dd r$ and $\dd t/\dd r$, integrating which we can compute the deflection angle and the travel time in section \ref{sec:dphianddt}. To this purpose, we first do some simple manipulations of the constraints/conditions
\eqref{eq:smunupnu}, \eqref{eq:pmupmu} and \eqref{eq:12smunusmunu}. Substituting the metric \eqref{eq:SASlinement} and using
\eqref{eq:zerocomps} in these equations, they become respectively
\begin{subequations}
\label{eq:sass}
\begin{align}
\lb \frac{B}{2} p^{t}+C p^{\varphi}\rb S^{t\varphi}+D p^{r} S^{tr}=&0,\label{eq:sass1} \\
\lb A p^{t}-\frac{B}{2} p^{\varphi}\rb S^{t \varphi}-D p^{r} S^{r\varphi }=&0,\label{eq:sass2} \\
\lb A p^{t}-\frac{B}{2}p^{\varphi}\rb S^{tr}+\lb \frac{B}{2} p^{t}+C p^{\varphi}\rb S^{r\varphi }=&0,\label{eq:sass3}
\end{align}
\end{subequations}
(the $\theta$-component equation is automatically satisfied)
\be
-A\lb p^{t}\rb^2+Bp^{t}p^{\varphi}+C\lb p^{\varphi}\rb^2+D\lb p^{r}\rb^2=-m^{2},\label{eq:sasm2}\ee
and
\begin{align}
-AD\lb S^{tr}\rb^2-\lb \frac{B^2}{4}+AC\rb \lb S^{t\varphi}\rb^2+CD\lb S^{r\varphi }\rb^2-BDS^{tr}S^{r\varphi }=J_m^{2}.\label{eq:sasj2}
\end{align}
Here and henceforth, $A,~B,~C,~D$ stand for the metric functions on the equatorial plane and therefore are only functions of $r$.
The equations \eqref{eq:sass} can be thought as a linear equation system of $S^{tr},~S^{t\varphi}$ and $S^{r\varphi}$, and one can check easily that the coefficient matrix of this system is rank 2. Consequently there is a redundant equation and we can choose any two of the three to use. Combining eqs. \eqref{eq:sass1}, \eqref{eq:sass2} and \eqref{eq:sasj2}, we can solve $S^{t r},~S^{t\varphi }$ and $S^{r\varphi }$ in terms of $p^t,~p^r$ and $p^\varphi$
\begin{subequations}
\label{eq:sassp}
\begin{align}
S^{t\varphi}=&-\alpha D p^{r},\label{eq:sassp1}\\
S^{tr}=&\alpha\left(Cp^\varphi+\frac{B}{2}p^t\right),\label{eq:sassp2}\\
S^{r\varphi }=&\alpha\left(\frac{B}{2}p^\varphi-Ap^t\right),\label{eq:sassp3}
\end{align}
\end{subequations}
where eq. \eqref{eq:sasm2}
is used to simplify the result and we introduced
\be
\label{fac}
\alpha=-\frac{s_jj }{\sqrt{D \left(B^2/4+AC\right)}}. \ee
Here we have defined the spin to mass ratio of the particle as
\be  j=\frac{J_m}{m}, \ee
and  $s_j=\pm$ is a sign that appears because we took the square root of eq. \eqref{eq:sasj2}.
In this work, without losing any generality, we will assume that the spin angular momentum of the spacetime (if any) is along the positive $z$ direction. Therefore, the $+$ (or $-$) sign in $s_j$ corresponds to the case that the particle's spin angular momentum is parallel (or antiparallel) to the  spacetime spin angular momentum.

Our next big step is to express the generalized momentum $p^\mu$ in terms of the metric functions, their derivatives and the coordinates. To do this, we notice that the stationary and axial symmetries of the spacetime \eqref{eq:SASlinement}  admit two Killing vectors
\begin{align}
\xi^{\lb 1\rb}_\mu=\lb 1,~0,~0,~0\rb, ~
\xi^{\lb 2\rb}_\mu=\lb 0,~0,~0,~1\rb.
\end{align}
Then according to eq. \eqref{eq:ck}, they generate two conserved quantities, the energy $E$ and total angular momentum $L$ parallel to spacetime spin, expressed in terms of $S^{tr},~S^{r\varphi }$ and $p^{\mu}$. Using eq. \eqref{eq:sassp}, the spin tensor components can be eliminated so we get
\begin{subequations}
\label{newel}
\begin{align}
E=&-\frac{1}{8}\left  \{\alpha\left[2\lb BA^\prime-AB^\prime\rb p^t+\lb 4CA^\prime+BB^\prime\rb p^{\varphi }\right] -8A p^t+4B p^{\varphi }\right \}, \label{eq:sasE}\\
L=&\frac{1}{8}\left \{\alpha \lsb-\lb 4AC^\prime+BB^\prime\rb p^t+2\lb BC^\prime-CB^\prime\rb p^{\varphi }\rsb+4B p^t+8C p^{\varphi }\right \},\label{eq:sasL}
\end{align}
\end{subequations}
where and henceforth $^\prime$ denotes the derivative with respect to the radial coordinate $r$.
Using eq. \eqref{newel} and the mass normalization equation \eqref{eq:sasm2}, the $p^t,~p^\varphi$ and $p^r$ can be solved in terms of the metric functions, their derivatives and other constants
\begin{subequations}
\label{eq:pmuinmetric}
\begin{align}
p^t=&-\frac{8\lb B\eta_1  +2C \eta_2  \rb}{\left(B^{2}+4 A C\right)\eta_3 },\label{newpt}\\
p^{\varphi}=&\frac{8\lb -2A\eta_1  +B \eta_2 \rb}{\left(B^{2}+4 A C\right)\eta_3 },\label{newpf}\\
\left(p^r\right)^2=&\frac1D\lsb \frac{64 \left(-A \eta_1^2+B  \eta_1\eta_2  +C  \eta_2^2\right)}{\left( B^2+4 A C\right)\eta_3^2}-m^2\rsb,\label{newpr}
\end{align}
\end{subequations}
where to simplify the notation we have defined
\begin{align}
\eta_1=&4L+\alpha \lb B^\prime L+2C^\prime E\rb,\\
\eta_2=&4E+\alpha \lb 2A^\prime L-B^\prime E\rb,\\
\eta_3=&-16+\alpha^2\left( B^{\prime 2}+4  A^\prime C^\prime\right).
\end{align}
The simpler version of these equations in the Schwarzschild spacetime was obtained in eqs. (101) to (103) of ref. \cite{Zalaquett:2013ifd}.

With the $S^{\mu\nu}$ obtained in eq. \eqref{eq:sassp} and $p^\mu$ in eq. \eqref{eq:pmuinmetric}, we can finally utilize the motion equation \eqref{eq:dotpmu}. The spin tensor in the right-hand side of this equation can be directly substituted by eq. \eqref{eq:sassp}.
The absolute derivative in the left-hand side can be computed using \eqref{absolutederivative} as
\be
\frac{\cdd p^\mu}{\dd \tau}\equiv\frac{\dd p^{\mu}}{\dd {\tau}}+\Gamma^\mu_{\rho\sigma} p^\rho u^\sigma\quad\quad (\mu=t,~r,~\varphi ).\label{eq:asdp}\ee
Here the Christoffel symbols $\Gamma^\mu_{\rho\sigma}$, as well as the Riemann tensors $R^\mu_{\nu\rho\sigma}$
in the right-hand side of eq. \eqref{eq:dotpmu}, can be readily expressed in terms of the metric functions in line element \eqref{eq:SASlinement}. The important point here is that: because all $p^t,~p^r$ and $p^\varphi$ are now functions of $r$ only as shown by eq. \eqref{eq:pmuinmetric} (note we are on the equatorial plane), the total derivative term in eq. \eqref{eq:asdp} when expressed using the chain rule, has only one component surviving for each $\mu$
\be \frac{\dd p^\mu}{\dd \tau}=\frac{ \partial p^\mu}{ \partial x^\nu}\frac{ \dd x^\nu}{\dd \tau }= \frac{ \partial p^\mu}{ \partial r}u^r \quad \quad (\mu=t,~r,~\varphi).\label{eq:dpdtau}\ee
Substituting eq. \eqref{eq:dpdtau} into \eqref{eq:asdp} and further into eq. \eqref{eq:dotpmu}, and then using eq. \eqref{eq:pmuinmetric}, we finally get rid of all spin tensor and generalized momentum, and obtain a system of three equations involving the metric functions, their derivatives, and the four velocity $u^\mu$

\begin{subequations}
\label{eq:eom1t}
\begin{align}
-\frac{1}{2}R^{t}_{\nu\rho\sigma}u^{\nu}S^{\rho\sigma}=&\frac{dp^t}{d\tau}+\Gamma^{t}_{\varphi r}\lb u^{\varphi}p^{r}+u^{r}p^{\varphi}\rb +\Gamma^{t}_{tr}\lb u^{t}p^{r}+u^{r}p^{t}\rb,\\
-\frac{1}{2}R^{\varphi}_{\nu\rho\sigma}u^{\nu}S^{\rho\sigma}=&\frac{dp^{\varphi}}{d\tau}+\Gamma^{\varphi}_{\varphi r}\lb u^{\varphi}p^{r}+u^{r}p^{\varphi}\rb +\Gamma^{\varphi}_{tr}\lb u^{t}p^{r}+u^{r}p^{t}\rb,\\
-\frac{1}{2}R^{r}_{\nu\rho\sigma}u^{\nu}S^{\rho\sigma}=&\frac{dp^r}{d\tau}+\Gamma^{r}_{rr}u^{r}p^{r}+\Gamma^{r}_{\varphi \varphi}u^{\varphi}p^{\varphi}+\Gamma^{r}_{tt}u^{t}p^{t}+\Gamma^{r}_{t\varphi}\lb u^{t}p^{\varphi}+u^{\varphi}p^{t}\rb,
\end{align}
\end{subequations}
where $S^{\mu\nu}$ are as in eq. \eqref{eq:sassp} and $p^\mu$ as in eq. \eqref{eq:pmuinmetric}.
This is indeed a linear and homogeneous equation system of $u^t,~u^r$ and $u^\varphi$. An explicit check shows that the coefficient matrix of this system is also rank 2, and therefore it allows us to solve $u^\varphi,~u^t$ in terms of $u^r$. The results are
\begin{align}
\frac{\dd\varphi}{\dd r}=&\frac{u^\varphi}{u^r}=\frac{\eta_7\lb A,C\rb\eta_1+8 \eta_2 \lcb \left[-16BD-8D\alpha'\left(B^2+4AC\right)\right]+\alpha  \eta_6 +\alpha^2  \eta_5 \rcb}{\eta_3 ^2 \left(B^2+4 AC\right) \sqrt{ D\lsb 64\left(-A \eta_1   ^2+B  \eta_1  \eta_2  +C  \eta_2  ^2\right)  \left(B^2+4 AC\right)^{-1}\eta_3 ^{-2}- m^2 \rsb}}, \label{eq:dphidrdef}\\
\frac{\dd t}{\dd r}=&\frac{u^t}{u^r}=\frac{\eta_7\lb C,A\rb\eta_2+8\eta_1\lcb \left[16BD-8D\alpha'\left(B^2+4AC\right)\right]+\alpha  \eta_6 -\alpha^2\eta_5 \rcb}{\eta_3 ^2 \left(B^2+4 AC\right) \sqrt{ D\lsb 64\left(-A \eta_1   ^2+B  \eta_1  \eta_2  +C  \eta_2  ^2\right)  \left(B^2+4 AC\right)^{-1}\eta_3 ^{-2}- m^2 \rsb}}, \label{eq:dtdrdef}
\end{align}
where
\begin{align}
&\eta_5=BD\lb B'^2+4A'C'\rb-DB'\lb B^2+4AC\rb^\prime-\lb B^2+4AC\rb \lb B'D'-2DB''\rb,\\
&\eta_6=-4\left[D\lb B^2+4AC\rb\right]^\prime,\\
&\eta_7\lb X,Y\rb=256 XD+8   \alpha^2\left[2D\left(4YX'^2+2BX'B'-XB'^2\right)\right.\nn\\
&~~~~~~~~~~~~~~~\left.  +2\left(B^2+4XY\right)\left(X'D'-2DX''\right)\right].
\end{align}
Eqs. \eqref{eq:dphidrdef} and \eqref{eq:dtdrdef} agree with eq. (46) of ref. \cite{Abdulxamidov:2022ofi} after some transformations.
Setting  $\alpha=0$, eqs. \eqref{eq:dphidrdef} and \eqref{eq:dtdrdef} reduce to eq. (3) of ref. \cite{Huang:2020trl} and eq. (7) of ref. \cite{Liu:2020mkf} respectively, which considered the case of spinless particles.

\section{Perturbative deflection angle and travel time \label{sec:dphianddt}}

\begin{figure}[htp!]
\centering
\includegraphics[width=0.6\textwidth]{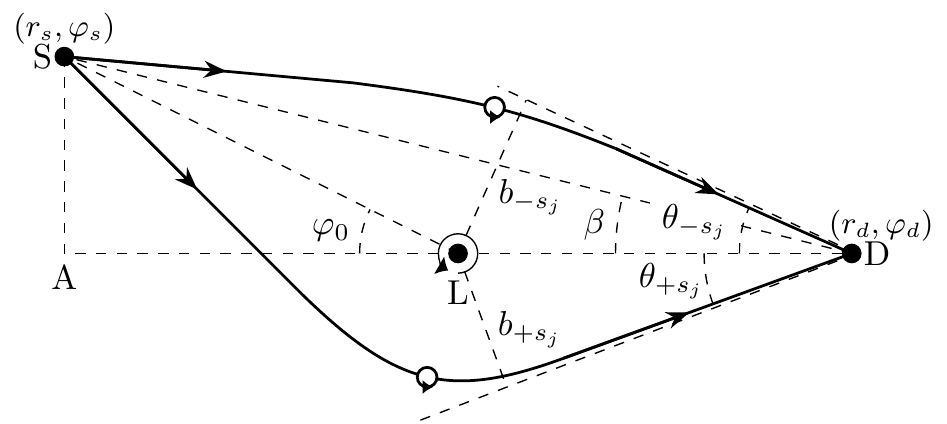}
\caption{The deflection and GL in a stationary and axisymmetric spacetime. S, L and D are the source, lens and detector respectively. $b_{+s_j}$ and $b_{-s_j}$ are the impact parameters from the counterclockwise and clockwise directions respectively (viewed from the top). The small circles on the path denote the particle spin $s_j$ ($+$ in the figure). $\theta_{\pm s_j}$ are the apparent angles of the images observed by D.}  \label{fig:lensing}
\end{figure}

We can now proceed to compute the deflection angle and the total travel time  for a particle propagating from the source at coordinates $(r_s,~\varphi_s)$ to the detector at $(r_d,~\varphi_d)$ (see figure \ref{fig:lensing}), by integrating  eqs. \eqref{eq:dphidrdef} and \eqref{eq:dtdrdef} respectively
\begin{align}
\Delta\varphi =&\lsb \int_{r_0}^{r_s}+\int_{r_0}^{r_d}\rsb \frac{\dd  \varphi}{\dd r}\dd r,\label{eq:dphidef}\\
\Delta t =&\lsb \int_{r_0}^{r_s}+\int_{r_0}^{r_d}\rsb \frac{\dd  t}{\dd r}\dd r,\label{eq:dtdef}
\end{align}
where $r_0$ is the minimal $r$ of the trajectory. We will only study the above quantities in the weak field limit, under which $r_s,r_d\gg r_0\gg M$ where $M$ is the characteristic mass of the spacetime. We will also assume that when $r \to+\infty$, the metric functions approach their asymptotic values in the following way, $A \to 1+\mathcal{O} \lb r^{-1}\rb$, $B \to \mathcal{O}\lb r^{-1}\rb$, $C \to r^2+\mathcal{O}\lb r^{1}\rb$, and  $D \to 1+\mathcal{O}\lb r^{-1}\rb$, which is true for asymptotically flat spacetimes. Then from eq. \eqref{newel}, $L$ and $E$ can be related to the particle's impact parameter $b$ and asymptotic velocity $v$ by
\begin{align}
E=&p^t\Big|_{r\to\infty}=\frac{m}{\sqrt{1-v^2}}, \label{eq:Erinfinity}\\
L=&\lb r^2 p^{\varphi}+s_jj p^t\rb\Big|_{r\to\infty}=\frac{s_o mvb}{\sqrt{1-v^2}}+\frac{s_jj m}{\sqrt{1-v^2}}.\label{eq:Lrinfinity}
\end{align}
Here we have used the definition of the orbital angular momentum $l$ of the particle
\be
r^2p^\varphi\Big|_{r\to\infty}\equiv l=\frac{s_o mvb}{\sqrt{1-v^2}}, 
\label{eq:oribtangdef}
\ee
where $s_o=\pm $ has two choices: $+$ (or $-$) corresponds to the case the particle's orbital angular momentum is parallel (or antiparrallel) to the spacetime spin angular momentum. Eqs. \eqref{eq:Erinfinity} and \eqref{eq:Lrinfinity} allow us to replace parameters $E$ and $L$ appearing in any equations by $b$ and $v$, including those in the integrands of eqs. \eqref{eq:dphidef} and \eqref{eq:dtdef}.

Usually, the integrals in eqs. \eqref{eq:dphidef} and \eqref{eq:dtdef}
can not be
carried out explicitly. Therefore previously we have developed perturbative techniques to compute these quantities for spinless particles \cite{Huang:2020trl,Liu:2020mkf}. In those works, a change of variables from $r$ to $u$ linked by a uniquely designed function (see ref. \cite{Huang:2020trl,Liu:2020mkf} for their exact forms)
\be p\lb\frac{1}{r}\rb=\frac{u}{b} \label{eq:pform}
\ee
was made to transform away the large $r_0$ in the integral limits and to make sure the  perturbatively expanded integrand is always integrable to obtain explicit series result. However, in the current case of spinning particles, the integrands in eqs. \eqref{eq:dphidef} and \eqref{eq:dtdef}
are significantly more complicated than the case of spinless particles. Therefore it also becomes very non-trivial to find a suitable change of variables that can satisfy the above two basic requirements.

One property of the above-mentioned function $p$
is that it satisfies a particularly simple relation with the apparent angles $\theta \lb r\rb$ (see figure \ref{fig:lensing}) as seen by a rest observer O located on the particle's trajectory at radius $r$:
\be \sin \theta \lb r\rb= b\cdot p\lb \frac{1}{r}\rb . \label{eq:pprop}
\ee
Therefore we can attempt to find the change of variables needed in this work also using the apparent angle.
Using the four-velocity of the observer O, i.e. $\left(1 / \sqrt{A\lb r\rb}, 0,0,0\right)$, one can find the projection of the four-velocity $u^\mu$ of the particle onto it
\begin{equation}
u^\mu_{\mathrm{proj}}= \frac{\sqrt{A}}{\lb 2 u^t A-u^\varphi B \rb} \left(\frac{B u^\varphi}{A}, 2u^r, 0, 2 u^\varphi \right).
\end{equation}
Further projecting this onto the radial direction yields the sine value of the trajectory's apparent angle at point O 
\begin{equation}
\label{eq:thetares}
\sin \theta\lb r\rb=\lsb 1+\frac{4AD}{B^{2}+4 A  C}\left(\frac{u^r}{u^\varphi}\right)^{2}\rsb^{-1/2}.
\end{equation}
Eqs. \eqref{eq:pform}, \eqref{eq:pprop} and \eqref{eq:thetares} hint us to first define a function $p\lb 1/r\rb$ in the same way as before
\begin{align}
p\lb \frac{1}{r}\rb \equiv \frac{\sin\theta\lb r\rb}{b}=\frac1b
\cdot \lsb 1+\frac{4AD}{B^{2}+4 A  C}\left(\frac{\dd r}{\dd \varphi}\right)^{2}\rsb^{-1/2}, \label{eq:pdefnew}
\end{align}
where $u^r/u^\varphi=\dd r/\dd\varphi$ is used and the latter is known in eq. \eqref{eq:dphidrdef}. Then the next step is to propose an analogous but much more complicated change of variables from $r$ to $u$ linked by the relation
\be
p\lb \frac{1}{r}\rb = \frac{u}{b}, \label{eq:covdef}
\ee
i.e.,
\be
\lsb 1+\frac{4AD}{B^{2}+4 A  C}\left(\frac{\dd r}{\dd \varphi}\right)^{2}\rsb^{-1/2}=u. \label{eq:rudirect}\ee
We note that the function $p\lb 1/r \rb$ in eq. \eqref{eq:pdefnew} will be immediately known once the metric functions are known.
Denoting the inverse function of $p \lb x\rb$ as $w \lb x\rb$, then clearly $r$ can be expressed in terms of $u$ using
\be
\frac{1}{r}=w \lb \frac{u}{b} \rb ~\text{or equivalently}~r=1/w\lb \frac{u}{b}\rb. \label{eq:rinu}
\ee

Now carrying out the change of variables in eq. \eqref{eq:covdef} or equivalently \eqref{eq:rinu}, the integral limits and the integrands of
eqs. \eqref{eq:dphidef} and \eqref{eq:dtdef}
become respectively
\begin{subequations} \label{eq:covtrans}
\begin{align}
&r_0\to 1,~r_{s,d}\to b\cdot p\lb \frac{1}{r_{s,d}}\rb \equiv \sin{\theta_{s,d}},\label{covtrans1}\\
&\dd r\to  -\frac{1}{p'_w \lb w\rb w^2}\frac1{b}\dd u,\label{covtrans2}\\
&\frac{\dd\varphi}{\dd r}\to \sqrt{\frac{4A \lb 1/w\rb D \lb 1/w\rb }{B \lb 1/w\rb ^2+4A \lb 1/w\rb C \lb 1/w\rb }}\frac{u}{\sqrt{1-u^2}} ,\label{covtrans3}\\
&\frac{\dd t}{\dd r}=\frac{\dd t}{\dd \varphi}\frac{\dd\varphi}{\dd r}\to \frac{t_{num} \lb r\to 1/w\rb}{\varphi_{num} \lb r\to 1/w\rb}\sqrt{\frac{4A \lb 1/w\rb D \lb 1/w\rb }{B \lb 1/w\rb ^2+4A \lb 1/w\rb C \lb 1/w\rb }}\frac{u}{\sqrt{1-u^2}} ,\label{covtrans4}
\end{align}
\end{subequations}
where $p'_w$ is the derivative of function $p \lb w\rb$ in \eqref{eq:pdefnew} with respect to its argument and $w = w \lb u/b\rb$.
In obtaining \eqref{covtrans1} and \eqref{covtrans3}, the definition of $r_0$
through $\dd r/\dd\varphi \big|_{r=r_0}=0$ and eq. \eqref{eq:rudirect} are used. The factor in front of $\dd u$ in eq. \eqref{covtrans2} is just the Jacobian factor for the change of variables. And in eq. \eqref{covtrans4}, $t_{num} (r)$ and $\varphi_{num} (r)$ are respectively the numerators of eqs. \eqref{eq:dphidrdef} and \eqref{eq:dtdrdef} as functions of $r$.
Collecting terms in eq. \eqref{eq:covtrans} together, $\Delta\varphi$ and $\Delta t$ become
\begin{align}
\Delta\varphi =&\lsb \int_{\sin\theta_s}^{1}+\int_{\sin\theta_d}^{1}\rsb y\lb \frac{u}{b} \rb \frac{\dd u}{\sqrt{1-u^2}}, \label{eq:phiinu}\\
\Delta t =&\lsb \int_{\sin\theta_s}^{1}+\int_{\sin\theta_d}^{1}\rsb z\lb \frac{u}{b} \rb \frac{\dd u}{u\sqrt{1-u^2}}, \label{eq:tinu}
\end{align}
where the integrands are
\begin{align}
y \lb\frac{u}{b}\rb =&\frac{1}{p'_w \lb w\rb w^2}\frac{u}{b}\sqrt{\frac{4A\lb 1/w\rb D\lb 1/w\rb }{B\lb 1/w\rb ^2+4A\lb 1/w\rb C\lb 1/w\rb }},\label{eq:ydefori}\\
z \lb\frac{u}{b}\rb =&
\frac{1}{p'_w \lb w\rb w^2}\frac{u^2}{b}\frac{t_{num}\lb 1/w\rb }{\varphi_{num}\lb 1/w\rb }
 \sqrt{\frac{4A\lb 1/w\rb D\lb 1/w\rb }{B\lb 1/w\rb ^2+4A\lb 1/w\rb C\lb 1/w\rb }}.\label{eq:zdefori}
\end{align}

The key point is to realize that in the weak field limit, these integrands can be expanded into series of small $u/b$,
\begin{align}
y\lb\frac{u}{b}\rb =&\sum_{n=0}^{} y_n \lb \frac{u}{b}\rb ^n, \label{eq:y}\\
z\lb\frac{u}{b}\rb =&\sum_{n=-1}^{} z_n \lb \frac{u}{b}\rb ^n,\label{eq:z}
\end{align}
whose coefficients $y_n$ and $z_n$ can be determined by the asymptotic behavior of the metric functions. In this work we will assume that the spacetime is asymptotically flat so that the asymptotic expansions of the metric functions take the following forms in the equatorial plane
\begin{align}
A\left(r\right) =\sum_{n=0}^\infty\frac{a_n}{r^n},\quad B\left(r\right)=\frac{1}{r} \sum_{n=0}^\infty\frac{b_n}{r^n},\quad C\left(r\right)=r^2\sum_{n=0}^\infty\frac{c_n}{r^n}, \quad
D\left(r\right)=\sum_{n=0}^\infty\frac{d_n}{r^n}.\label{eq:ABCD}
\end{align}
In asymptotically Minkowski spacetime,  one should have $a_0=c_0=d_0=1$ and we can always identify $a_1=-2M$, i.e. $-2$ times the ADM mass $M$ of the spacetime.

Substituting the above into eq. \eqref{eq:pdefnew}, we can work out the asymptotic expansion of $p \lb 1/r\rb$ as
\begin{align}
\label{seriesp}
p \lb \frac1r \rb =\frac{1}{r}-\lcb \frac{c_1}{2}-\frac{a_1}{2 v^{2}}+\frac{s_o\lsb  b_0- s_j j \lb a_1  +d_1\rb \rsb}{2 b v}\rcb\frac{1}{r^2}   +\mathcal O\lb \frac{1}{r^3} \rb.
\end{align}
This can be inverted using the Lagrange Inversion Theorem to find the series expansion of the inverse function $w\lb u/b\rb$
\begin{align}
\label{seriesw}
w \lb\frac{u}{b} \rb=\frac{u}{b}+ \lcb \frac{c_1}{2}-\frac{a_1}{2 v^{2}}+\frac{s_o\lsb  b_0- s_j j \lb a_1  +d_1\rb \rsb}{2 b v}\rcb\frac{u^2}{b^2}  +\mathcal O \lb \frac{u^3}{b^3}\rb .
\end{align}
Substituting expansions \eqref{seriesp} and \eqref{seriesw}, and eq. \eqref{eq:ABCD} into
eqs. \eqref{eq:ydefori} and \eqref{eq:zdefori}, we can work out their expansion coefficients $y_n$ in eq. \eqref{eq:y} and $z_n$ in \eqref{eq:z}. The first few of them are
\begin{subequations}
\label{eq:ynexp}
\begin{align}
y_0 =&s_o,\label{eq:ynexp1}\\
y_1=&\frac{s_o d_1}{2}-\frac{s_o a_1}{2v^2}+\frac{b_0}{2bv}-\frac{s_j j\lb a_1+d_1 \rb}{2bv},\label{eq:ynexp2}\\
y_2=&-\frac{s_o}{8}  \left(c_1-d_1\right)^2+\frac{s_o}{2} \left(c_2+d_2\right)-\frac{s_o}{2v^2} \left[a_1 \left(c_1+d_1\right)-2 a_1^2+2 a_2\right]-\frac{a_1 b_0}{bv^3}\nn\\
&+\frac{3s_o b_0^2}{4b^2v^2}+\frac{-2a_1 b_0+b_0 \lb c_1+d_1 \rb +2b_1}{2bv}+\frac{s_j j a_1 \lb a_1 +d_1 \rb}{bv^3}\nn\\&+\frac{s_j j \lsb 3a_1^2-2a_1 c_1 -4a_2 + \lb c_1-d_1 \rb^2 -4 \lb c_2+d_2 \rb \rsb}{4bv}\nn\\&+\frac{3s_o j \lb a_1 + d_1 \rb \lsb  a_1+d_1 -2s_j b_0 +j \lb a_1 +d_1 \rb \rsb}{4b^2 v^2},\label{eq:ynexp3}\\
y_3=&\frac{3 s_j j b  a_1  }{2 v}+\mathcal{O}\lb  M^3 \rb,\label{eq:ynexp4}
\end{align}
\end{subequations}
and
\begin{subequations}
\label{eq:timedelay}
\begin{align}
z_{-1}=&\frac{1}{v},\label{eq:timedelay1}\\
z_{0}=&\frac{a_1 \lb 1-2v^2 \rb+d_1v^2  }{2v^3},\label{eq:timedelay2}\\
z_1=&\frac{8 a_1^2-4 a_1 \lb c_1+d_1\rb -8 a_2-\lb c_1-d_1\rb^2+4 c_2+4 d_2}{8 v}+\frac{s_o b_0  \lb -4 a_1+c_1+d_1\rb +2 s_o b_1 }{4 b v^2} \nn\\
&+\frac{b_0^2}{4b^2 v^3}+\frac{s_o s_j j}{8 b v^2}\lsb 7a_1^2-2a_1 \lb c_1-2d_1\rb -4 \lb a_2 +c_2 \rb +\lb c_1 +d_1 \rb^2-d_2 \rsb\nn\\&-\frac{ s_j j b_0 \lb a_1+d_1 \rb }{2b^2 v^3}+\frac{j^2 \lb a_1+d_1\rb ^2 }{4 b^2 v^3}
,\label{eq:timedelay3}\\
z_2=&\frac{s_o b b_0 }{2}+\frac{ s_o s_j j b a_1}{2} \lb 1 +\frac{1}{v^2} \rb+ \mathcal{O}\lb M^2 \rb.\label{eq:timedelay4}
\end{align}
\end{subequations}
Setting $j=0$ in eq. \eqref{eq:ynexp}  reduces it to its counterpart for spinless particles, which appeared in eq. (4.8)  of ref. \cite{Huang:2020trl}. Further setting the coefficients $b_n$ to zero, eq. (2.23) of ref. \cite{Xu:2021rld} (after setting $\hat{q}=0$ and $a_{0n}=0$), i.e., the result in static and spherically symmetric spacetime, is obtained.
Setting $j=0$ in eq. \eqref{eq:timedelay} reduces it to eqs. (26) to (29) of ref. \cite{Liu:2020mkf} for spinless signals. Further setting $b_n$ to zero, we recover eq. (2.24) (after setting $\hat{q}=0$ and $a_{0n}=0$) of ref. \cite{Xu:2021rld}.

With $y_n$ and $z_n$ known in eqs. \eqref{eq:phiinu} and \eqref{eq:tinu}, it is clear then $\Delta\varphi$ and $\Delta t$ can be written as a series of integrals of the form
\be
I_n \lb \theta_i \rb =\int_{\sin \theta_i }^1\frac{u^n}{\sqrt{1-u^2}}\dd u
~~ \lb i=s,d;~ n=-2,-1,\cdots \rb. \label{I}
\ee
These integrals can be easily worked out and the results are triangular functions of $\theta_i$, whose closed forms are given in eq. \eqref{eq:inres} in appendix \ref{sec:app1}.
Therefore finally we obtain the series forms of $\Delta\varphi$ and $\Delta t$
\begin{align}
\Delta\varphi  =&\sum_{i=s,d}\sum_{n=0}^\infty y_n\frac{I_n \lb \theta_i \rb}{b^n},\label{eq:dphifinal}\\
\Delta t =&\sum_{i=s,d}\sum_{n=-1}^\infty z_n\frac{I_{n-1} \lb \theta_i \rb}{b^n}.\label{eq:dtfinal}
\end{align}

For the infinite source and detector distance case, using the definition of $\theta_{s,d}$ in eq. \eqref{eq:thetares} and noticing the asymptotics \eqref{eq:ABCD}, we can verify that $\theta_i(r\to\infty)=0~\lb i=s,d\rb$. Consequently, from eq. \eqref{eq:inreslim} we see that $I_n~(n\geq0)$ become independent of $b$ and both $\Delta\varphi$ and $\Delta t$ become pure power series of $\lb M/b\rb$. For the finite $r_{s,d}$ case however, $\theta_i$ will be nonzero and then $I_n\lb \theta_i\rb$ have a weak and non-power functional dependence on $r_i$. Therefore $\Delta\varphi$ and $\Delta t$ are only quasi-power series of $\lb M/b\rb$ in this case. What is more useful in this situation are the expansions of $I_n\lb\theta_i\rb$ in the small quantities $\lb M/b\rb$ and $\lb b/r_i\rb$, as given in eq. \eqref{eq:saslnexp} in appendix \ref{sec:app1}.  These expansions enable us to write $\Delta\varphi$ and $\Delta t$ as a dual power series of $\lb M/b\rb$ and $\lb b/r_i\rb$
\begin{align}
\Delta \varphi=&\sum_{i=s,d}s_o\lsb\frac{\pi  }{2}+\frac{1 }{2 b}\left(d_1-\frac{a_1}{v^2}\right)+\frac{1}{2 b^2}\bigg\{\frac{\pi  }{16}  \Big[4 \left(c_2+d_2\right)-\left(c_1-d_1\right)^2\right]+\frac{s_o b_0}{v}+\frac{s_j j \lb a_1-d_1\rb}{ v}\nn\\
&\left.\left.+\frac{\pi   \left[a_1 \left(2 a_1-c_1-d_1\right)-2 a_2\right]}{4 v^2} \rcb-\frac{b }{r_i}+\mathcal{O}\lb \epsilon^3\rb\rsb ,\label{aformula}
\\
\Delta t=&\sum_{i=s,d} \lsb \frac{r_i}{v}+ \frac{1}{2v}\lb c_1-\frac{a_1}{v^2}\rb -\frac{b^2}{2vr_i}+ \ln \left(\frac{2 r_i}{b}\right)\cdot \frac{\lb d_1-2 a_1 \rb v^2+a_1}{2 v^3}+\frac{1}{16b} \bigg\{ \pi\Big[ 8 a_1^2-8 a_2\right.\nn\\&\left.\left.- 4 a_1 \lb c_1+d_1\rb-\lb c_1-d_1\rb^2+4\lb c_2+ d_2\rb\rsb+ 8s_o b_0 \lb 1+\frac{1}{v^2} \rb+8s_o s_j  j \lb a_1-\frac{d_1}{v^2}\rb\rcb  \nn\\
&+\mathcal{O}\lb r_i\epsilon^4,~j\epsilon^2\rb\bigg] ,\label{eq:ttgenfth}
\end{align}
where and henceforth $\epsilon$ stands for either infinitesimal $\lb M/b\rb$ or $\lb b/r_i\rb$. 
Note in eq. \eqref{aformula}, there is naturally no $\lb M/r_i\rb^1$ or $\lb b/r_i\rb^2$ terms at the $\epsilon^2$ order. Eq. \eqref{aformula} for spinless particles in the Kerr spacetime was obtained in eq. (2.18) of ref. \cite{Huang:2020trl}, and for spinless but charged particles was obtained in the Reissner-Nordstr\"{o}m (RN) spacetime in eq. (2.30) of ref. \cite{Xu:2021rld}. While the time delay eq. \eqref{eq:ttgenfth} for spinless particles was obtained in eq. (35) of ref. \cite{Liu:2020mkf}. Therefore our work not only generalized them to arbitrary stationary and axisymmetric spacetimes but also included the effect of particle spin.

\section{GL of particles with spin\label{sec:glgeneral}}

In this section, we will assume that the test particles are the particles common in astrophysics and having nonzero spin, such as neutrinos, other leptons or small cosmic ray particles. We will first establish an exact GL equation and solve the impact parameters allowing particles to reach the detector. Then using these impact parameters, the apparent angles of the images and the time delays between them will be solved.

Using the deflection angle with finite distance effect taken into account, it is simple to establish a GL equation without invoking any further geometrical or physical approximations beyond the fact that $\Delta\varphi$ is computed perturbatively. This GL equation is nothing but the very definition of $\Delta\varphi$ itself (refer to figure \ref{fig:lensing})
\be
\Delta\varphi \lb b\rb= s_o\pi+\varphi_0. \label{eq:gleqdef}\ee
Here $\varphi_0$ is the angle of the source's radial direction against the lens-detector axis. It can also be replaced by another commonly used angle $\beta$, i.e., the angular separation of the source from the lens if the lens did not exist. From figure \ref{fig:lensing}, $\varphi_0$ and $\beta$ are connected by the geometrical relation
\be r_s\sin\varphi_0=\lb r_d+r_s\cos \varphi_0\rb\tan\beta. \label{eq:betaphi0rel}\ee
Later on, in plots in section \ref{sec:apps} we will use freely either $\varphi_0$ or $\beta$.

Since the deflection $\Delta\varphi \lb b\rb$ in eq. \eqref{aformula} is a rational function of $b$, eq. \eqref{eq:gleqdef} actually can be recast into a polynomial of $b$. In order for this equation to be solvable, here we will use the $\Delta\varphi$ truncated only to the order of $\lb M/b\rb^2$ and $\lb b/r_{s,d}\rb^2$. Using eq. \eqref{aformula}, eq. \eqref{eq:gleqdef} becomes
\begin{align}
 \pi + s_o\varphi_0 =\pi+\lb d_1-\frac{a_1}{v^2} \rb \frac{1}{b}-b\lb \frac{1}{r_s}+\frac{1}{r_d}\rb+\frac{c_{s_os_j}}{b^2},\label{GL equation}
\end{align}
where $c_{s_os_j}$ is the coefficient of $1/b^2$ in eq. \eqref{aformula},  
\begin{align}
c_{s_os_j}=&\frac{1}{2} \lcb\frac{\pi   \left[a_1 \left(2 a_1-c_1-d_1\right)-2 a_2\right]}{4 v^2}-\frac{\pi  }{16} \left[\left(c_1-d_1\right)^2-4 \left(c_2+d_2\right)\right]+\frac{s_o b_0}{v}\rcb\nn\\&+\frac{s_o s_j j }{2 v}\left(a_1-d_1\right).\label{Cco}
\end{align}
The last term of eq. \eqref{GL equation} is much smaller than others and therefore can be thought as a perturbation. We include this term because, as we will show next, it is through this term that the spacetime spin and particle spin can influence the solution of the impact parameters and then the apparent angles of the images. Eq. \eqref{GL equation} can be solved to find solutions $b_{s_os_j}$
\begin{align}
b_{s_os_j}= b_{0s_o}+b_{1s_os_j}+\mathcal{O}\lb b_{1s_os_j}^2/b_{0s_o}\rb,\label{bpm}
\end{align}
where
\begin{align}
b_{0s_o}=&\frac{\varphi_0 r_d r_s}{2 \lb r_d+r_s \rb} \lb \sqrt{1+\eta}-s_o \rb, \label{eq:b0sodef}\\
b_{1s_os_j}=&\frac{c_{s_os_j}}{\left[\lb a_1/v^2-d_1\rb+  3b_{0s_o}^2\lb  1/r_s+1/r_d \rb+2s_o b_{0s_o} \varphi_0\right]},\label{correction}\\
\eta=&\frac{4\lb r_d +r_s \rb}{\varphi^2_0 r_d r_s v^2} \lb -a_1+d_1 v^2 \rb.\label{eta}
\end{align}
Clearly, it is through the correction term $b_{1s_os_j}$, which contains $c_{s_os_j}$ as a factor, that the impact parameters become dependent on the spacetime spin $b_0$ and particle spin $j$. Without this, $b_{s_os_j}$ would be exactly the same as the impact parameters in ordinary spherically symmetric spacetime for particles without spin at all (see eq. (39) of ref. \cite{Liu:2020mkf}). It will be useful to note by a simple order estimation that the size of $b_{1s_os_j}$ is of order $\mathcal{O}\lb M\rb$.

With the impact parameters known, using eq. \eqref{eq:pprop} then we can find the apparent angles of the images seen by the detector at $r_d$
\begin{align}
\theta_{s_os_j}=\sin ^{-1}\left[b_{s_os_j} \cdot p\left( \frac{1}{r_{d}}\right)\right]=\frac{b_{0s_o}}{r_d}+\frac{b_{1s_os_j}}{r_d}
+\mathcal{O}\lb \epsilon^3\rb,\label{expandangle}
\end{align}
where in the second step the expansion of function $p\lb x\rb$ and $\theta_i$ in eq. \eqref{seriestheta} are used, and we keep the result to the second order of $\epsilon$, i.e., the first non-trivial order containing $j$. It is seen that at this order, the correction from the particle spin $j$ comes from its correction to the impact parameters $b_{1s_os_j}$ in the GL equation, but not the higher orders in the expansion (see eq. \eqref{seriestheta}) of the function $p\lb x\rb$ in the definition \eqref{eq:pprop} of $\theta$. In section \ref{sec:apps} we will analyze the effect of $j$ in more detail in particular spacetimes.

Substituting $b_{s_os_j}$ in eq. \eqref{bpm} into the total travel time \eqref{eq:ttgenfth} and subtracting the cases with different $s_o$ and/or $s_j$, we are able to get the time delays between different images. For signals with same spin but from different sides of the lens, their time delay $\Delta^2 t_o$ is
\begin{align}\label{timedelaycase1}
\Delta^2 t_o \equiv & \Delta t \lb b_{+s_j} \rb - \Delta t \lb b_{-s_j} \rb \nn\\
=& \frac{1}{2v}\lb\frac{1}{r_s}+\frac{1}{r_d}\rb \lsb \lb b_{0-}^2 -b_{0+}^2  \rb+2\lb b_{0-}b_{1-s_j}-b_{0+}b_{1+s_j}\rb\rsb\nn\\&+\frac{1}{2v}\lb \frac{a_1}{v^2}-2a_1+d_1 \rb \lsb \ln\lb\frac{b_{0-}}{b_{0+}}\rb+\lb\frac{b_{1-s_j}}{b_{0-}}-\frac{b_{1+s_j}}{b_{0+}}\rb\rsb\nn\\&-\frac{\pi}{8v} \lsb 8 a_1^2 -4  a_1 \lb c_1+d_1 \rb -8 a_2- \lb c_1+d_1 \rb ^2+4 \lb c_2+d_2\rb \rsb \lb\frac{1}{b_{0-}}-\frac{1}{b_{0+}}\rb\nn\\
&+b_0\lb 1+\frac{1}{v^2}\rb  \lb\frac{1}{b_{0-}}+\frac{1}{b_{0+}} \rb+s_j j\lb a_1-\frac{d_1}{v^2}\rb \lb \frac{1}{b_{0-}}+\frac{1}{b_{0+}}\rb + \mathcal{O}\lb r_i\epsilon^4,~j\epsilon^2 \rb.
\end{align}
After setting $j$ to zero for spinless particles, this agrees with eq. (43) of ref. \cite{Liu:2020mkf}. If we substitute the RN metric and $j=0$, it further reduces to eq. (3.11) of ref. \cite{Xu:2021rld} (after setting the particle charge to zero there). Note that the third term is always much smaller than the first and second terms (see ref. \cite{Liu:2020mkf}) and therefore in the specific spacetimes considered in section \ref{sec:apps} we will directly ignore it.

We can also use the same eqs. \eqref{bpm} and \eqref{eq:ttgenfth} to compute the time delay $\Delta^2 t_{j_1j_2}$ between a signal with spin $j_1$, velocity $v+\Delta v$ and energy $E_1$ and a signal with spin $j_2$, velocity $v$ and energy $E_2$, but from the same side of the lens
\begin{align}
\label{timedelay2}
\Delta^2 t_{j_1j_2}\equiv & \Delta t \lb b_{s_os_j1}\rb  - \Delta t \lb b_{s_os_j2}\rb  \nn\\
=& \lcb b_{0 s_o} \lb  \frac{1}{r_s}+\frac{1}{r_d} \rb+\frac{1}{2b_{0 s_o}}\lsb d_1-a_1 \lb 2-\frac{1}{v^2}\rb\rsb\rcb  \frac{\lb b_{1 s_o s_j2}-b_{1 s_o s_j1}\rb}{v}\nn\\&+\frac{s_o \lb s_{j2}j_2-s_{j1}j_1\rb \lb d_1- a_1 v^2 \rb}{b_{0s_o}v^2}+\frac{\lb r_s+r_d \rb}{2v^2} \lb \frac{m_1^2}{E_1^2 }
-\frac{m_2^2}{E_2^2 }\rb  +\mathcal{O}\lb r_i\epsilon^4,~j\epsilon^2,~\Delta v^2\rb,
\end{align}
where only the first two non-trivial orders are kept. The second to last term in this equation is due to the difference in signal velocity and therefore proportional to $\Delta v$. It originates from the first term on the right-hand side of eq. \eqref{eq:ttgenfth} and contains no general relativistic effect at this order. This term agrees with the result in ref. \cite{Zatsepin:1968kt,Jia:2017oar} and can be shown to be much smaller than other terms in most application situations involving supernova neutrinos. Therefore in most applications in section \ref{sec:apps} we will directly ignore it.

We note that the time delay \eqref{timedelay2} is of order $\mathcal{O}\lb Mj/b  \rb$ and its size depends crucially on the value of $j$. It is this time delay that differs significantly
from the ones in previous literature \cite{Liu:2020mkf,Xu:2021rld}, which did not take into account the effect of particle spin on the time delay. Therefore this time delay might be  used to deduce information about the signal's spin to mass ratio.
In section \ref{sec:apps} we will study this time delay for neutrinos and show that it might be measurable by neutrino observatories if $j$ is large enough.
Finally let us also point out that
if one is interested in the time delays between signals with different spins and meanwhile from different sides of the lens, then a combination of eqs. \eqref{timedelaycase1} and \eqref{timedelay2} will yield the desired result.

\section{Application to typical spacetimes \label{sec:apps}}

In this section, we will apply the general formalism and results in previous sections to particular stationary and axisymmetric spacetimes. In subsection \ref{subsec:kerr}, special attention will be paid to the Kerr spacetime case for its importance in astronomy. We will also use the Kerr spacetime results to check the correctness of the perturbative method, including the deflection angle, apparent angles and time delays found in section \ref{sec:dphianddt}. Furthermore, effects of various parameters of the spacetime and particle (especially the particle spin) on the above quantities will be studied.
In subsection \ref{subsec:teo}, we will apply the results to the deflection and GL of spinning particles in the Teo wormhole spacetime.

\subsection{Kerr spacetime results\label{subsec:kerr}}
In Kerr spacetime, the metric functions on the equatorial plane are given by \cite{Misner:1973prb}
\begin{align}
A\left(r\right)= 1-\frac{2M}{r},\quad
B\left(r\right)=-\frac{4aM}{r},\quad
C\left(r\right)= r^2+a^2 +\frac{2Ma^2}{r},\quad
D\left(r\right)= \frac{r^2}{r^2-2Mr+a^2},\label{Kerr}
\end{align}
where $M$ and $a=J_K/M$ are respectively the total mass and angular momentum per unit mass of the central body.
Their asymptotic expansion coefficients are easily found to be
\begin{subequations}
\label{eq:Kerr}
\begin{align}
&a_0=1,~a_1=-2M,~a_{n\geq 2}=0,\\
&b_0=-4aM,~b_{n\geq 1}=0,\\
&c_0=1,~c_2=a^2,~c_3=2a^2M,~c_1=c_{n\geq 4}=0,\\
&d_0=1,~d_1=2M,~d_2=4M^2-a^2,~\cdots.
\end{align}
\end{subequations}

\subsubsection{Deflection angle and apparent angles}

Substituting eq. \eqref{eq:Kerr} into \eqref{eq:ynexp}, the first few $y_n$ coefficients of the deflection angle in Kerr spacetime are obtained
\begin{subequations}
\label{yKinKerr}
\begin{align}
y_{K,0}=&s_o,\\
y_{K,1}=& M\lsb s_o\lb1+\frac{1}{v^2}\rb-\frac{2 a}{bv}\rsb,\\
y_{K,2}=&M^2\left[3s_o  \lb \frac{1}{2}+\frac{2}{v^2}\rb-\frac{4  a}{b v}\lb 3+\frac{2}{v^2}\rb+\frac{12 s_o a^2}{b^2 v^2}\right],\\
y_{K,3}=& M^3 \lsb \frac{s_o }{2} \left(5+\frac{45}{v^2}+\frac{15}{v^4}-\frac{1}{ v^6}\right) -\frac{9 a }{bv}\left(5+ \frac{10}{v^2}+\frac{1}{v^4}\right)+\frac{24 s_o   a^2}{b^2v^2} \left(5+\frac{3}{v^2}\right)-\frac{80a^3  }{b^3 v^3}\rsb \nn\\
&
+Ma^2\lsb \frac{3 s_o}{2} \left(1+\frac{1}{v^2}\right)-\frac{3a}{b v}\rsb- 3s_j  Mj\lsb \frac{ b}{v}-  s_o a \left(1+\frac{1}{v^2}\right)+ \frac{ a^2 }{bv}\rsb\nn\\&-12 M j^2\lb s_o+\frac{ a }{bv }\rb, \label{eq:examp}
\end{align}
\end{subequations}
and substituting these into eq. \eqref{eq:dphifinal}, the deflection angle in Kerr spacetime's equatorial plane becomes
\be
\Delta\varphi_K=\sum_{i=s,d}\sum_{n=0}^\infty y_{K,n}\frac{I_n \lb \theta_i \rb}{b^n}.
\label{eq:dphifinalk}\\
\ee
In the large $r_{s,d}$ limit, as previously stated, this can be transformed to a dual power series of $\lb M/b\rb$ and $\lb b/r_{s,d}\rb$. Substituting the coefficients \eqref{eq:Kerr} directly into eq. \eqref{aformula}, $\Delta\varphi_K$ is transformed to
\begin{align}
\label{eq:dfk}
\Delta\varphi _{K}=&\sum_{i=s,d}s_o\left[\frac{\pi }{2}+\left(1+\frac{1}{v^2}\right)\left(\frac{M}{b}\right)-\frac{b}{r_{i}}\right.\nn\\&\left.+\left(\frac{3 \pi}{8}+\frac{3 \pi}{2 v^2}-\frac{2s_o\hat{a}}{v}-\frac{2s_o s_j \hat{j}}{v}\right)\left(\frac{M}{b}\right)^2+\mathcal{O} \left(\epsilon^3\right)\right],
\end{align}
where we have introduced $\hat{j}=j/M$ and $\hat{a}=a/M$. For even higher order terms containing $\hat{j}^{n\geq 2}, ~\hat{a}^{n\geq 2}$ and $\hat{a}\hat{j}$ coupling, one can refer to eq. \eqref{eq:dphihexp} in appendix \ref{sec:highphik}. 

\begin{figure}[htp!]
\centering
\includegraphics[width=0.6\textwidth]{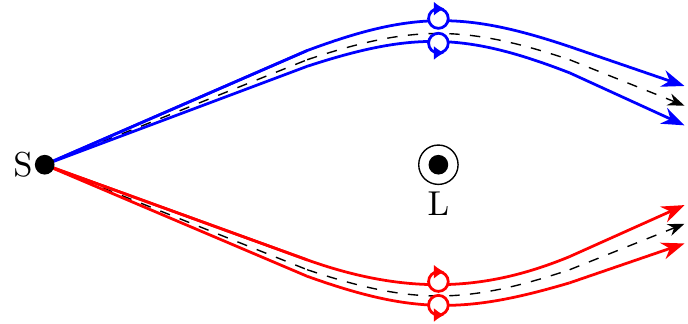}
\caption{The effect of the signal spin on its deflection angle. Small circles, S and L are respectively the signal particles, source and lens. Arrows on the small circles and the trajectories (solid curves) represent the direction of the spin and orbital angular momenta of the signal respectively. The black dashed lines represent the path of spinless signals. The spin of the central lens can be in either direction or zero. }
\label{fig:jtodphi}
\end{figure}

Eq. \eqref{eq:dfk} allows us to analyze clearly the effect of signal spin on the deflection angle.
For spinless particles, setting $\hat{j}=0$, this formula agrees with eq. (5.2) of ref. \cite{Huang:2020trl} (after setting the charge $\hat{q}$ to zero there).
Comparing to the spinless case, the effect of nonzero $j$ starts to manifest from the second order, i.e., the $\mathcal{O}\lb M/b\rb^2$ order. This is similar to the effect of the spacetime spin \cite{Huang:2020trl} or the magnetic dipole on a charged signal \cite{Li:2022cpu}. Apparently, the sign of this term depends on the signs of $s_o$ and $s_j$. When the orbiting direction of the signal is anti-clockwise so that $s_o=+$, a signal with positive (or negative) spin $s_j=+$ (or $s_j=-$) will decrease (or increase) the size of the deflection angle. In contrast, when the orbiting direction of the signal is clockwise so that $s_o=-$, a signal with positive (or negative) spin $s_j=+$ (or $s_j=-$) clearly will increase (or decrease) the size of the deflection angle. In summary, the size of the deflection angle is reduced if the orbital and spin angular momenta of the signal are parallel, and enlarged if antiparallel. We illustrate these effects schematically in figure \ref{fig:jtodphi}.

Qualitatively, this effect of spin on the deflection angle is consistent with the effective force due to spin-orbit coupling in gravity found in Wald's work \cite{Wald:1972sz}. Various other effects due to spin of test particles in gravity also suggest the same extra effective force compared to Newtonian gravity.
For example, Pomeransky et al. \cite{Pomeransky:2000pb} found the acceleration of the spinning particle in a gravitational field, from which we can also assert the existence of the same extra force. Hojman and Asenjo \cite{Hojman:2018evi} found that when the spin and orbital angular momenta are parallel (or antiparallel), then there will exist an extra repulsive (or attractive) effect which allows smaller (or larger) tangential velocity of circular motions. Zhang et al. \cite{Zhang:2017nhl}, Zhang and Liu \cite{Zhang:2018eau} and Shaymatov et al. \cite{Shaymatov:2021nff} found that the particle spin can affect its inner-most stable circular orbit parameters in a way depending on the relative signs of spin and orbital or total angular momenta.
The same spin-orbital direction dependence is also found in its effect on the angular velocities and momenta of general circular orbits \cite{Mohseni:2010rm}.

When $r_{s,d}$ is infinite, eq. \eqref{eq:dfk} is further simplified to
\begin{align}
\label{final1}
\Delta\varphi_K=s_o\left\{\pi +2\left(1+\frac{1}{v^2}\right)\left(\frac{M}{b}\right)+\left(\frac{3 \pi}{4}+\frac{3 \pi}{ v^2}-\frac{4s_o\hat{a}}{v}-\frac{4s_o s_j \hat{j}}{v}\right)\left(\frac{M}{b}\right)^2+\mathcal{O}\lb  \epsilon^3\rb\right\}.
\end{align}
The order $\mathcal{O}\lb \hat{j}M^2/b^2\rb$ term in this equation for ultra-relativistic particles (setting $v=1$)  reduces to the corresponding term in
refs. \cite{Barrabes:2003ey,Vines:2017hyw,Bini:2018zxp,Liu:2021zxr}.
The $\mathcal{O}\lb \hat{j}M^n/b^n\rb ~(n=3,4)$ terms in ref. \cite{Bini:2018zxp} also agree with our eq. \eqref{eq:dphihexp} for $v=1$. 
However, the higher order terms involving $\hat{j}^{n\geq 2}$ in ref. \cite{Vines:2017hyw,Bini:2018zxp,Liu:2021zxr} are different from ours (see eq. \eqref{eq:dphihexp}) because of their omission of some spin terms at higher order.

\begin{figure}[htp!]
\centering
\includegraphics[width=0.6\textwidth]{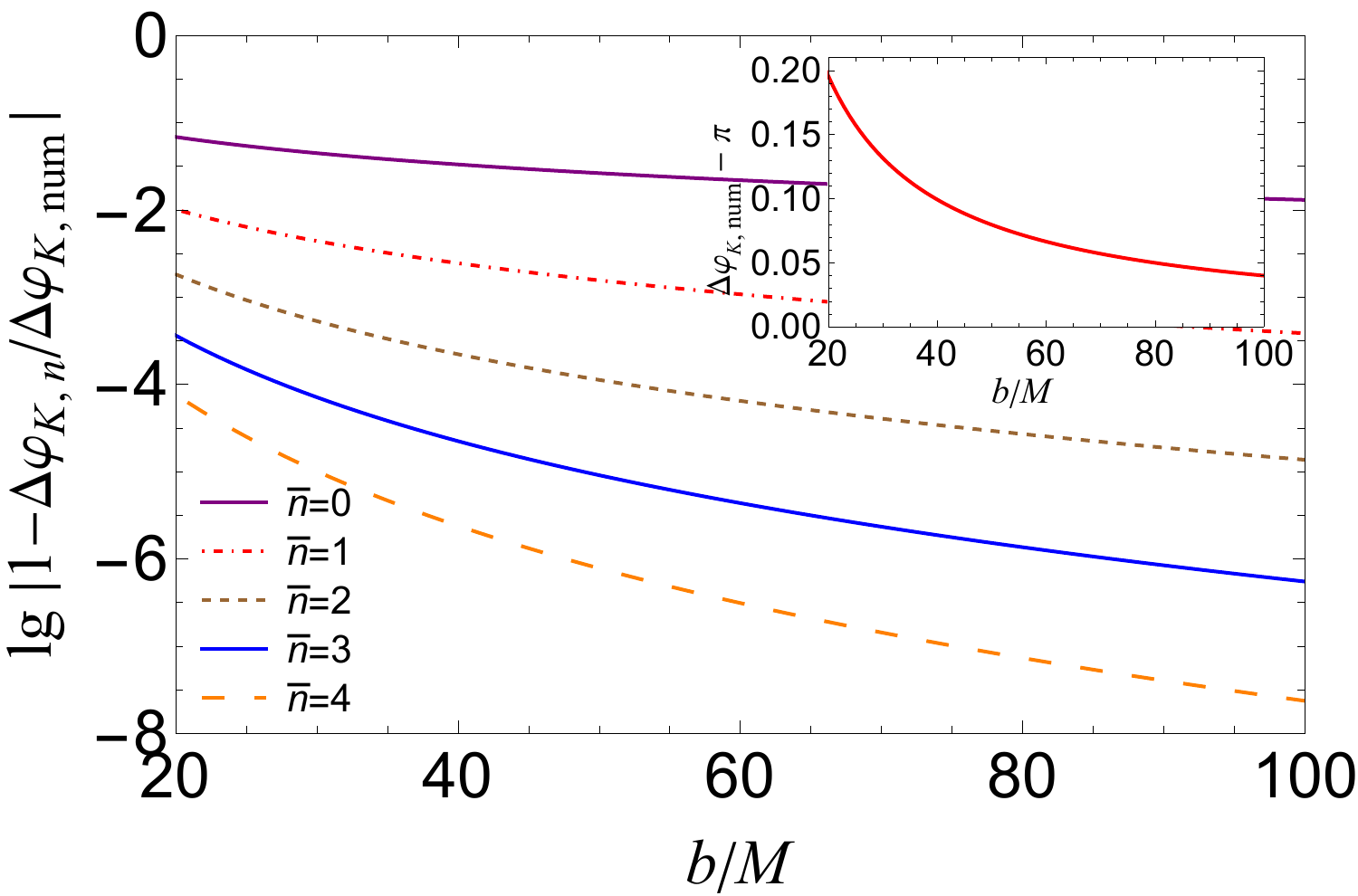}
\caption{The difference between $\Delta\varphi_{K,\bar{n}}$ and $\Delta\varphi_{K,\mathrm{num}}$ as a function of $b$. We choose $\hat{a}=1/2,~\hat{j}=1/200,~s_o=s_j=+,v=0.99$ and $r_s=r_d=10^6M$.
\label{fig:comp1}  }
\end{figure}

To examine the validity of our general formalism and the perturbative deflection result, eq. \eqref{eq:dphifinalk}, we can construct a truncated deflection angle
\be \Delta\varphi_{K,\bar{n}}\equiv \sum_{i=s,d}\sum_{n=0}^{\bar{n}} y_{K,n}\frac{I_n \lb \theta_i \rb}{b^n},
\label{eq:dphiktrunc}
\ee
where $\bar{n}$ is the truncation order, and compare it with the numerical deflection $\Delta\varphi_{K,\mathrm{num}}$ obtained by direct numerical integration of definition \eqref{eq:dphidef}. The numerical integration can achieve very high accuracy and therefore can be thought as the true deflection angle. In figure \ref{fig:comp1} we plot the difference between $\Delta\varphi_{K,\bar{n}}$ and $\Delta\varphi_{K,\mathrm{num}}$ as a function of the impact parameter $b$ with other parameters fixed. It is seen that as the truncation order increases, the perturbative result approaches the numerical value roughly exponentially, to less than $\sim 10^{-7}$ for the fifth order $(\bar{n}=4)$ result. The larger the impact parameter $b$, the smaller the deflection (seen from the inset) and the more accurate the perturbative result, as expected from a quasi-inverse power series form like eq. \eqref{eq:dphiktrunc}. Note that even when $b$ is as small as $20M$, at which point the deflection actually reaches 0.2 [rad] -- not a small angle anymore, the fifth order perturbative deflection is still very close to the true deflection.

Having verified the correctness of the perturbative $\Delta\varphi_K$ in eq. \eqref{eq:dfk}, we can now use it to study the effect of the spacetime parameters $(M,~\hat{a})$, the particle's intrinsic parameter $j$ and kinetic parameters ($s_o,~s_j,~b,~v)$ on the deflection angle. Among these, the effects of parameters $(M,~\hat{a},~b,~v)$ have been well studied in previous works \cite{Liu:2015zou,Jia:2020xbc,Huang:2020trl,Liu:2020mkf}. In this work, we will mainly focus on the effect of $s_j$ and $j$. A few observations can be immediately made from eq. \eqref{eq:dfk}.
First we note that the change of the angular coordinate at the lowest order is $s_o\pi$. This agrees with our previous setup that the spacetime spin is along the $+ z$ direction and $s_o=+$ corresponds to counterclockwise motion of the particle. Secondly, we note that both the spacetime spin and the particle spin contribute to the deflection angle starting from the order $\lb M/b\rb^2$. Thirdly, we note that the proportional constant of the contribution of the spacetime spin is $|\hat{a}/v|$, while that of the particle spin is $|\hat{j}/v|$. If the test object is of macroscopic size and its angular momentum $J_m$ is solely due to its self-rotation (including Kerr BH), so that $J_m$ does not exceed the angular momentum of an extreme Kerr BH of the same mass, then we would expect that $j\leq m$. Consequently $\hat{j}=j/M\leq m/M\ll 1$ while $\hat{a}=a/M$ could reach order 1.
Therefore, for these test objects the contribution of the spacetime spin to the deflection angle is usually much larger than that of the particle spin.

However, if we apply this to high energy astroparticles, then their $\hat{j}$ might be comparable to or larger than unity.
One particular interesting example is the extragalactic neutrino whose spin is $\hbar/2$ and mass is extremely small. Denoting the mass of the neutrino mass eigenstate $|\nu_i\rangle ~(i=1,2,3)$ as $m_i$, then it is easy to verify that its $\hat{j}_i$ is
\be
\hat{j}_i=\frac{\hbar/2}{m_i M}=0.668\cdot \frac{10^{-10}\text{~[eV]}}{m_i} \cdot\frac{M_\odot}{M}. \ee
Clearly from eq. \eqref{eq:dfk}, for a solar mass lens, if $m_i<10^{-11}$ [eV], then the particle's spin contribution to the deflection angle will be larger than other contributions at the same order, including that of the lens spin.
For other heavier particles such as protons, neutrons or even electrons however, their $\hat{j}$ are usually very small. In table \ref{tab:1} we have computed the value of $j$ and $\hat{j}$ for protons, electrons as well as three neutrino masses and an extreme Kerr BH of solar mass. The latter might be considered a test particle when it is orbiting a much heavier lens such as the Sgr A* SMBH.

\begin{table*}[htp!]
\renewcommand\arraystretch{1.2}
\centering
\resizebox{\textwidth}{!}{
\begin{tabular}{p{24mm}||p{22mm}p{22mm}p{22mm}p{22mm}p{22mm}|p{25mm}}
\hline\hline
 & Proton & Electron  & $|\nu\rangle$ mass 1  & $|\nu\rangle$ mass 2 &$|\nu\rangle$ mass 3 & Extr. Kerr BH \\
\hline
 $m~\mathrm{[eV]}$ & $9.38\times 10^8$   & $5.11\times 10^5 $  & $1.00 \times 10^{-4}$ & $1.00 \times 10^{-8}$ & $1.00 \times 10^{-16}$ & $1.00 \times  \mathrm{M_\odot}$ \\
 $j/m$ & $8.47 \times 10^{37}$ & $2.85 \times 10^{44}$&$7.45 \times 10^{63}$&$7.45 \times 10^{71}$&$7.45 \times 10^{87}$&1.00\\
$\hat{j}_1=j/M_\odot$ & $7.12\times10^{-20}$ & $1.31\times10^{-16}$ & $6.68\times10^{-7}$ & $6.68\times10^{-3}$ &$6.68\times 10^5$& ---  \\
$\hat{j}_2=j/M_\mathrm{Sgr A^*}$ & $1.74\times10^{-26}$ & $3.19\times10^{-23}$ & $1.63\times10^{-13}$ & $1.63\times10^{-9}$ &$1.63\times10^{-1}$&$2.44 \times 10^{-7}$ \\
\hline\hline
\end{tabular}}
\caption{The value of $j,~\hat{j}_1$ and $\hat{j}_2$ for protons, electrons and neutrinos of different masses, as well as a solar mass extreme Kerr BH. For $\hat{j}_1$ and $\hat{j}_2$, we choose the lens mass to be $M_\odot$ and mass of the Sgr A* SMBH ($4.1\times 10^6 \mathrm{M_\odot}$ \cite{Andreas:2018}) respectively.}
\label{tab:1}
\end{table*}

In observation, it is however not the deflection angle $\Delta\varphi_K$ but the apparent angles that are directly observable. Therefore substituting the expansion coefficients \eqref{eq:Kerr} into the impact parameter \eqref{bpm} and then into eqs. \eqref{eq:b0sodef} to \eqref{expandangle}, we find the apparent angles in the Kerr spacetime to be
\be
\theta_{K,s_o s_j}=\frac{b_{K,0s_o}}{r_d}+ \frac{b_{K,1s_os_j}}{r_d}
+\mathcal{O}\lb \epsilon^3\rb,
\label{eq:thetakerr}
\ee
where
\begin{subequations}
\label{eq:kbkdefs}
\begin{align}
b_{K,0s_o}=&\frac{\varphi_0 r_d r_s}{2\lb r_d+r_s\rb}\lb \sqrt{ 1+\eta_K} -s_o \rb,\\
b_{K, 1 s_o s_j}=&\frac{\eta_K\lsb -16 s_o \lb a+ s_j j\rb v+3M \pi \lb 4+v^2 \rb \rsb}{32\lb 1+v^2\rb\sqrt{
1+ \eta_K}\lb \sqrt{
1+ \eta_K}-s_o\rb} ,\label{eq:kb1}\\
\eta_K=&\frac{8M\lb r_d +r_s \rb}{\varphi^2_0 r_d r_s } \lb 1+ \frac{1}{v^2} \rb. \label{eq:kbkdefs3}
\end{align}
\end{subequations}
We see from eq. \eqref{eq:thetakerr} that both the spacetime spin $a$ and particle spin $j$ affect the apparent angles only in the second order through $b_{K,1s_os_j}$. The square bracket in the numerator of $b_{K,1s_os_j}$ implies that (only) when $\hat{a}$ and/or $\hat{j}$ are comparable to 1, their effects become significant in this order. However, only when $\hat{j}\gtrsim b_{K,0s_o}/M\gg 1$, the effect of particle spin might be significant in the overall apparent angle.

\begin{figure}[htp!]
\centering
\includegraphics[width=0.6\textwidth]{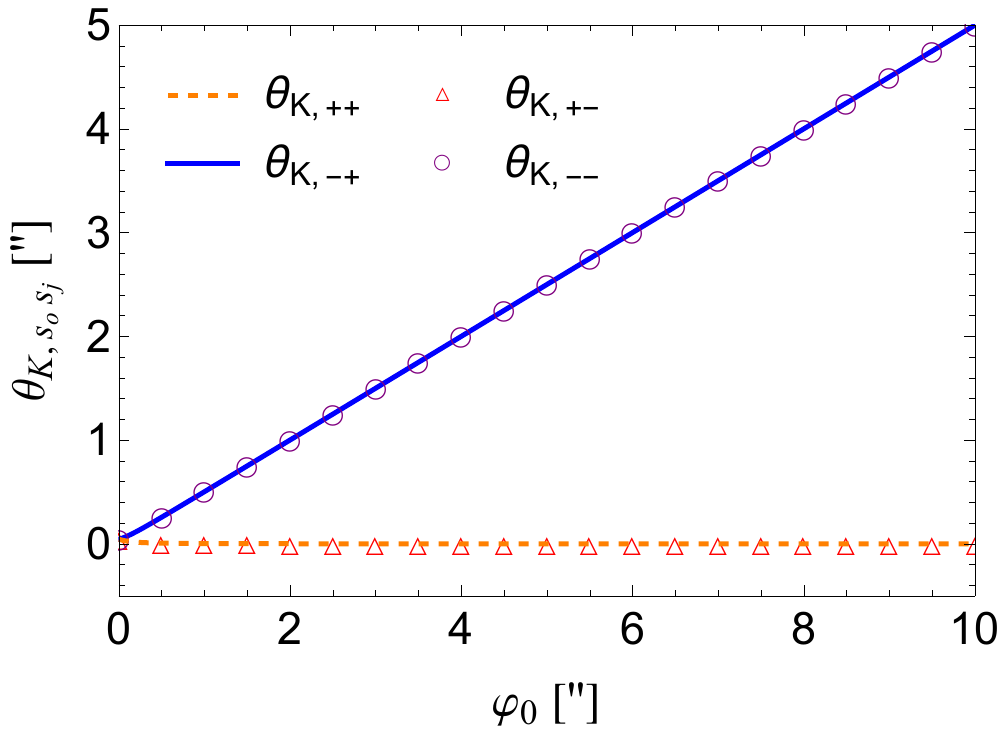}\\
(a)\\
\centering
\includegraphics[width=0.6\textwidth]{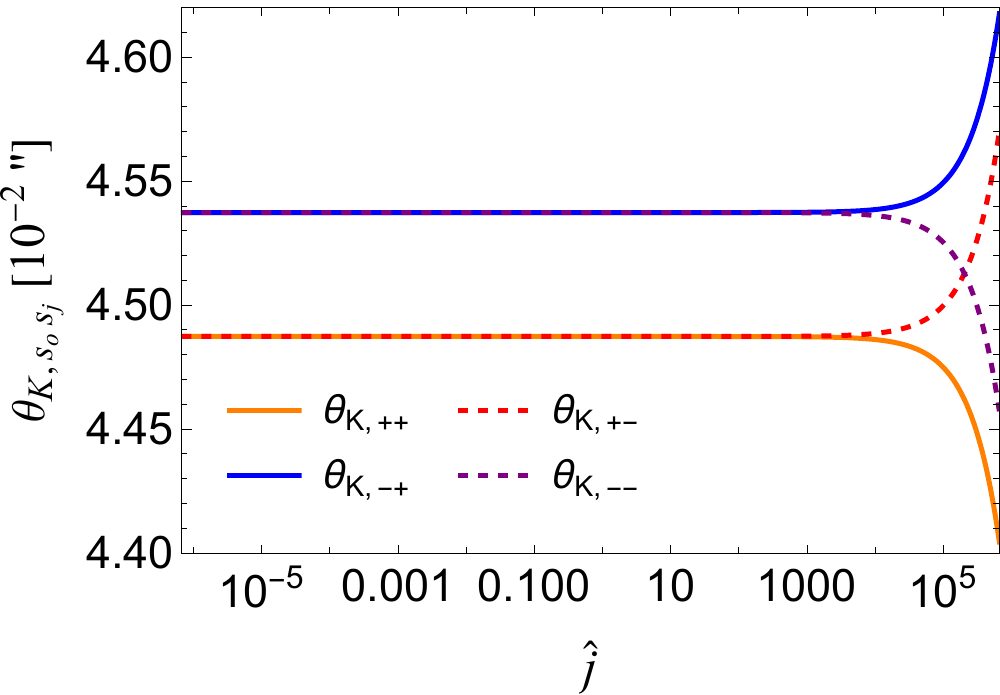}\\
(b)
\caption{The apparent angle \eqref{eq:thetakerr} as functions of $\varphi_0$ (a) and $\hat{j}$ (b) for a solar mass lens with $\hat{a}=0.1$ and $r_d=r_s=2~\mathrm{[pc]}$ as the lens. We use a medium value $\hat{j}_1=6.68 \times 10^{-3}$ in table \ref{tab:1} in (a) and $\varphi_0=10^{-3}~[^{\prime\prime}]$  in (b). In (b), we varied $\hat{j}$ from $6.68 \times 10^{-7}$ to $6.68 \times 10^{5}$ according to the third row of table \ref{tab:1}. In both plots,
$E_\nu=10~[\mathrm{MeV}]$ is used to deduce the signal's velocity.}
\label{fig:thetakerrplot}
\end{figure}

In figure \ref{fig:thetakerrplot}, we plot $\theta_{K,s_os_j}$ as functions of the source position $\varphi_0$ and particle spin $\hat{j}$ using a solar mass lens located at $2~\mathrm{[pc]}$ as the lens.
It is seen from figure \ref{fig:thetakerrplot} (a) for most range of $\varphi_0$ considered,  the image of a clockwise orbit ($s_o=-$) has a larger apparent angle than that of an anti-clockwise orbit ($s_o=+$). This is consistent with our definition of $\varphi_0$ as shown in figure \ref{fig:lensing} and agrees with former analysis \cite{Xu:2021rld}.
For the chosen $\hat{a}$ and (small) $\hat{j}$ in this plot, changing spacetime and/or signal spin directions (effectively $s_o$ and $s_j$) will not have any observable effect to the curves, consistent with the fact that these quantities appear only in the second order of deflection angle.
From figure \ref{fig:thetakerrplot} (b), it is seen that for the given small $\varphi_0=10^{-3}~[^{\prime\prime}]$, when $\hat{j}$ is smaller than $10^3$, which is roughly the size of $b_{K,0s_o}/M$ in this system, the apparent angles are only sensitive to the direction $s_o$ of the spacetime spin. If $\hat{j}$ exceeds this value, the apparent angles start to deviate from their values of spinless particles. Moreover, for the images with same $s_o$, the one with a $s_j=-s_o$ will have a larger apparent angle. This is consistent with the observation after eq. \eqref{eq:dfk} that signals with antiparallel (or parallel) spin and orbital angular momenta will have a larger (or smaller) deflection angle. When $\hat j$ reaches the maximum value of $1.63\times 10^5$ considered in table \ref{tab:1}, the apparent angle can change by $0.2\times 10^{-2} ~[^{\prime\prime}]$.
From eq. \eqref{eq:thetakerr} we see that the physical parameters that determine the apparent angles are $(r_s,~r_d,~M,~\varphi_0,~a,~j)$.
In observation, usually the distance $r_s,~r_d$ as well as the lens mass $M$ can be determined by other conventional means, therefore if the apparent angles of these four images in figure \ref{fig:thetakerrplot} (b) with different $s_o$ and $s_j$ are all measured and distinguished, the value of $j$ and $a$ will be well constrained. For neutrinos, a well constrained $j$ implies their masses $m=j\hbar/2$ can be constrained too.

\begin{figure}[htp!]
\centering
\includegraphics[width=0.6\textwidth]{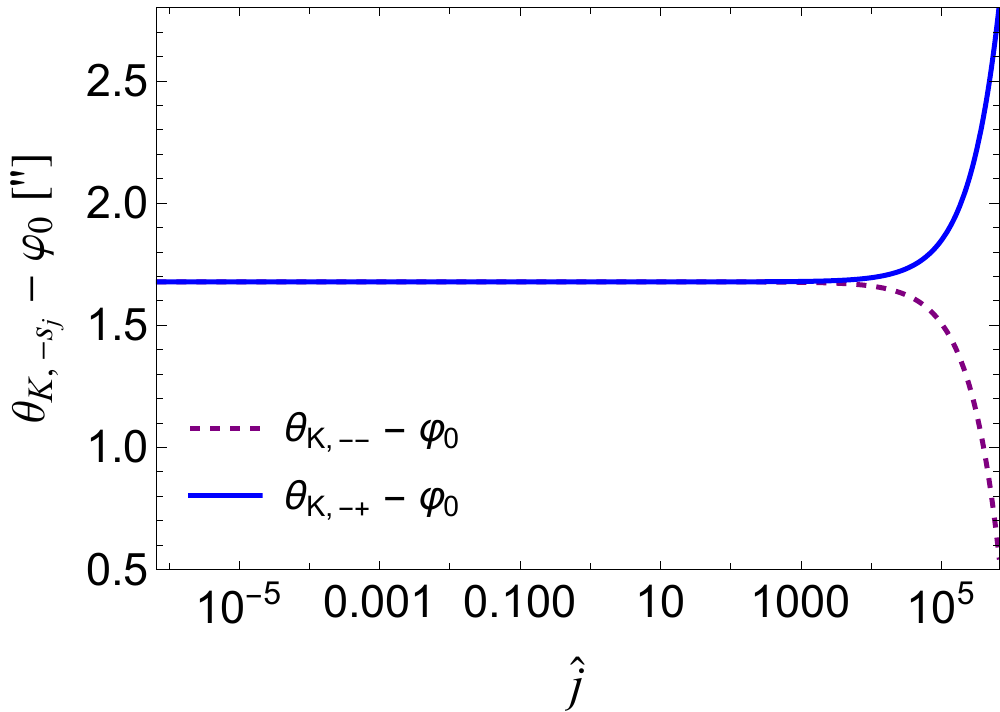}
\caption{The apparent angle \eqref{eq:thetakerr} as a function of $\hat{j}$ using the Sun as the lens. Physical value of $r_d=1~\mathrm{[AU]},~\hat{a}=0.218$ \cite{Iorio:2011pc} are used. For other parameters, we assume $r_s=8.1~\mathrm{[kpc]}$ and $\varphi_0= 1.04\theta_\odot$. }
\label{fig:thetakerrplot2}
\end{figure}

We also attempted to plot the apparent angles using the Sun as the lens to see the effect of the signal spin. However, since the Sun itself has a very large angular radius $\theta_\odot\approx 15.98 ~[^\prime]$ against us, the corresponding $\varphi_0$ will be large too. In this case, two of the images with $s_o=+$ will be very weakly lensed so that their apparent angles will be within the solar radius $\theta_\odot$ and not observable. In figure \ref{fig:thetakerrplot2}
therefore we only plot the apparent angles $\theta_{K,-s_j}$ for the same range of $\hat j$ as in figure \ref{fig:thetakerrplot} for a fixed $\varphi_0=1.04\theta_\odot$. It is seen that similar to the case in figure \ref{fig:thetakerrplot} (b), when $\hat j\gtrsim 10^3$, the apparent angles of two spin directions start to deviate from each other. When $\hat j$ reaches about $10^5$, they can differ by about $2.3~[^{\prime\prime}]$. This difference is about 1000 times larger than that in figure \ref{fig:thetakerrplot} (b) and therefore easier to detect from an observational point of view.

\subsubsection{Total travel time and time delays}

To find the total travel time of spinning signals in the Kerr spacetime, we first substitute the coefficients \eqref{eq:Kerr} into \eqref{eq:timedelay} to find the first few $z_n$
\begin{subequations}
\begin{align}
\label{eq:ext}
z_{K,-1}=&\frac{1}{v},\\
z_{K,0}=&M \lb \frac{3}{v}-\frac{1}{v^3}\rb,\\
z_{K,1}=&M^2 \lb \frac{15}{2v} -\frac{10s_o a}{bv^2}+\frac{4a^2}{b^2v^3}\rb,\\
z_{K,2}=&M^3\lsb \frac{1}{2v}\lb 35+\frac{35}{v^2} -\frac{7}{v^4}+\frac{1}{v^6}\rb -\frac{2 s_o a}{bv^2} \lb 35+\frac{14}{v^2}-\frac{1}{v^4}  \rb +\frac{12a^2}{b^2 v^3}\lb 7+\frac{1}{v^2} \rb -\frac{32s_o a^3}{b^3 v^4}  \rsb \nn\\&+M \lsb \frac{a^2}{2v}\lb 5+\frac{1}{v^2} \rb -2s_o a\lb b +\frac{a^2}{bv^2}\rb +\frac{s_j a j}{v} \lb 3+\frac{1}{v^2} \rb  - b s_o s_j j \lb 1+\frac{1}{v^2} \rb \right.\nn\\
&\left. -\frac{8j}{v}\lb j-\frac{s_o a}{bv}\rb -\frac{2s_o s_j a^2 j}{bv^2}\rsb.
\end{align}
\end{subequations}
If the coefficients were directly substituted into eq. \eqref{eq:ttgenfth},
the $\Delta t$ in the large  $r_{s,d}$ limit becomes
\begin{align}
\label{az}
\Delta t_K=&\sum_{i=s,d} \lcb\frac{r_i}{v}+\frac{M}{v^{3}}-\frac{1}{2 v}\frac{b^2}{r_i}+\left[\frac{15 \pi}{4 v }-s_o\left(2 \hat{a}+ s_j \hat{j}\right)\left(1+\frac{1}{v^{2} }\right)\right]\frac{M^2}{b}\right.\nn\\&\left.+\frac{M}{v}\left(3-\frac{1}{v^2}\right) \ln \lb \frac{2 r_{i}}{b} \rb+\mathcal{O}\lb r_i\epsilon^4,~j\epsilon^2\rb\rcb.
\end{align}

For the time delay purpose however, it is easier to directly substitute
eq. \eqref{eq:Kerr} into eqs. \eqref{timedelaycase1} and \eqref{timedelay2} to find the two fundamental time delays
\begin{align}
\Delta^2 t_{K,o} = & \ln \lb \frac{\sqrt{\eta_K+1}+1 }{\sqrt{\eta_K+1}-1 }\rb \cdot \frac{2 M \left(3 v^2-1\right) }{v^3}+\frac{\sqrt{\eta_K+1}}{\eta_K}\cdot \frac{4 M \lb 1+v^2 \rb}{ v^3}-a \varphi_0 \sqrt{\eta_K+1}\nn\\
&+\frac{3\varphi_{0} M\pi\lb 1-3v^2\rb\lb 4+v^2\rb}{32v\lb 1+v^2\rb^2}-\lb a+s_j j\rb\varphi_0 \lsb\frac{v^2\sqrt{\eta_K+1}}{1+v^2}\right.\nn\\&\left.-\frac{\lb v^2-3\rb \lb \sqrt{\eta_K+1}+1 \rb}{4\lb 1+v^2\rb^2}\rsb+ \mathcal{O}\lb r_i \epsilon^4,~ j\epsilon^2 \rb, \label{timedelayinkerr}\\
\Delta^2 t_{K,j_1j_2}=&\frac{\lb s_{j2} j_2-s_{j1} j_1\rb\varphi_0}{2}\lb s_o\sqrt{1+\eta_K}+1\rb\lcb\frac{v^2}{1+v^2}-\frac{\lb v^2-3\rb\lb\sqrt{1+\eta_K}+s_o\rb}{4\lb1+v^2\rb^2\sqrt{1+\eta_K}}\rcb\nn\\&+\mathcal{O}\lb r_i\epsilon^4,~j\epsilon^2\rb,
\label{eq:kdtsj}
\end{align}
where $\eta_K$ was given in eq. \eqref{eq:kbkdefs3} and $\varphi_0$ can be interchanged with $\beta$ using eq. \eqref{eq:betaphi0rel}. Eq. \eqref{timedelayinkerr} after setting $j$ to zero reduces to eq. (53) of ref. \cite{Liu:2020mkf} and (4.9) of \cite{Xu:2021rld} (after setting charges to zero there).

\begin{figure}[htp!]
\centering
\includegraphics[width=0.6\textwidth]{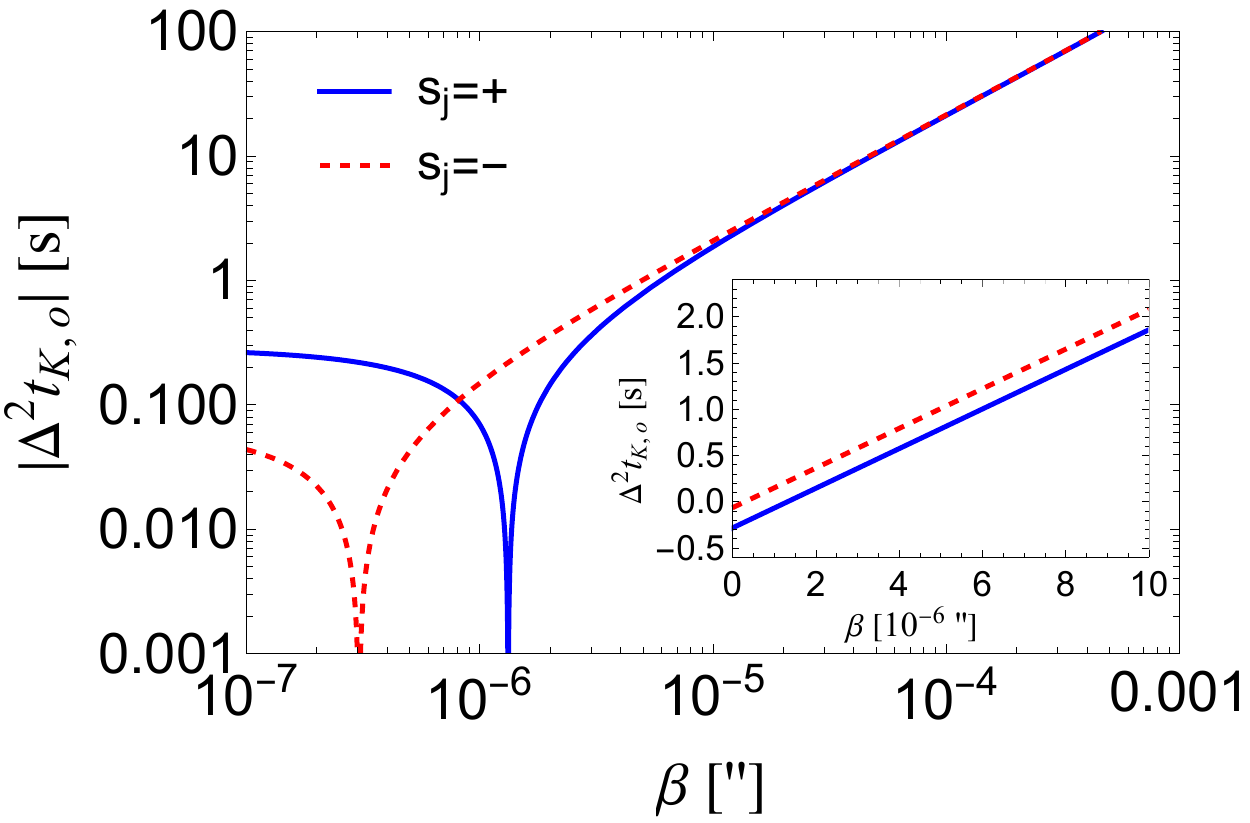}\\
(a)\\
\centering
\includegraphics[width=0.6\textwidth]{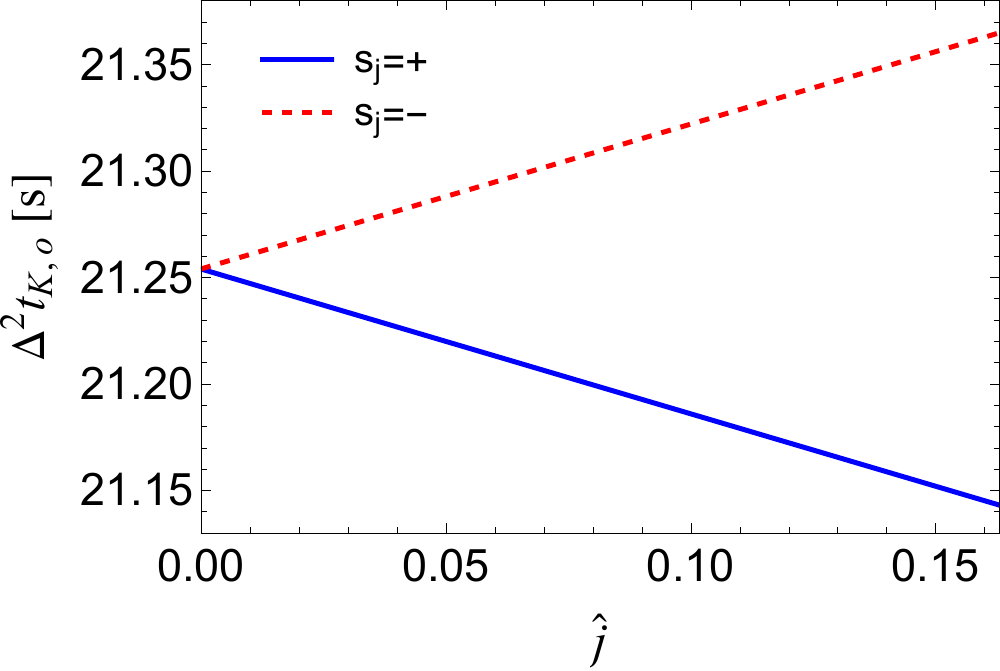}\\
(b)
\caption{The time delay \eqref{timedelayinkerr} as functions of $\beta$ (a) and particle spin $\hat{j}$ (b). We used the Sgr A* SMBH as the lens and assumed $r_s=r_{S_{39}}=1.113 \times 10 ^{-3}$ [pc], $r_d=8.1$ [kpc], $\hat{a}=0.1$ \cite{Fragione:2020khu}. Other parameters are: $\hat{j}=1.63 \times 10^{-1}$ in (a) and $\beta=10^{-4}$ [$^{\prime\prime}$] in (b). In (b), we varied $\hat{j}$ to $1.63 \times 10^{-1}$ according to the fourth row of table \ref{tab:1}. In both plots,
$E_\nu=10$ [MeV] and the $|\nu\rangle$ mass 3 in table \ref{tab:1} are used to deduce the velocity of the signal.}
\label{fig:tdkerrplot}
\end{figure}

It was known in ref. \cite{Liu:2020mkf} that for spinless signals, when the source angle $\varphi_0$ is small, the contribution of spacetime spin $a$ to time delay $\Delta^2 t_{K,o}$ will dominate, and therefore measuring $\Delta^2 t_{K,o}$ can be used to constrain $a$. Now for signals with spin, we see from eq. \eqref{timedelayinkerr} that for $j$ to have a relatively larger contribution to $\Delta^2 t_{K,o}$, we will have to study its large $\eta_K$ limit, while keeping $r_d/M$ and $\hat{j}$ large.
In figure \ref{fig:tdkerrplot} we plot $\Delta^2 t_{K,o}$ as functions of $\beta$ and $\hat j$ by assuming the Sgr A* SMBH is the lens. To compare with ref. \cite{Liu:2020mkf}, we will assume that the source is also at the radius of an S star, namely S$_{39}$ \cite{2017ApJ...837...30G}. From figure \ref{fig:tdkerrplot} (a) it is seen that for $\beta\lesssim  10^{-6}~[^{\prime\prime}]$, the third and fifth terms of eq. \eqref{timedelayinkerr} are larger than other terms combined so that the time delay in this range of $\beta$ is negative. That is, due to the effect of $a$ and $j$, the total travel time for a counter-clockwise path is smaller than a clockwise one, even for some positive $\beta$. As $\beta$ increases, the time delay increases almost linearly for both spin directions, reaching $\sim 2.1\times 10 ~\mathrm{[s]}$ at $\beta\approx 10^{-4}~[^{\prime\prime}]$. In addition, we also see that the time delay for signals with spin parallel to $a$ ($s_j=+$) is separated from that of spin antiparallel to $a$ ($s_j=-$) by about $0.2\sim 0.3$ [s], which is much larger than the time resolution of typical neutrino observatories ($10^{-9}$ [s]) and the characteristic timescale of certain neutrino signals (e.g. the neutronization peak in the supernova neutrino spectrum lasts about $\sim 10^{-2}$ [s] \cite{Jia:2017oar}). Note that this difference is independent of the value of $\hat{a}$.

Figure \ref{fig:tdkerrplot} (b) shows the dependence of $\Delta^2 t_{K,o}$ on $\hat{j}$. We choose $\beta=10^{-4}~[^{\prime\prime}]$ in the plot so that $\eta_K\approx 2.25\times 10^2\gg 1$.
Indeed, from eq. \eqref{timedelayinkerr} we see that for $\eta_K\gg 1$
\begin{align}
\Delta^2 t_{K,o} \lb v \to 1,~\eta_K\gg 1 \rb\approx\text{const.}-\frac{13a+5s_jj}{2}\sqrt{\frac{M\lb r_s+r_d \rb}{r_s r_d}}+ \mathcal{O}\lb r_i\epsilon^4,~j\epsilon^2\rb.  \label{eq:dt1kerrlim}
\end{align}
Then clearly the slopes of the curves in figure \ref{fig:tdkerrplot} (b) are determined by the coefficient of $j$, which is independent of the values of $a$ or $\beta$ as long as $\eta_K\gg 1$. In other words, even for Schwarzschild spacetime, the time delay $\Delta t^2_{K,o}$ in the $\eta_K\gg 1$ case is linear to $j=J_m/m$ of the signal with slope $5s_j\sqrt{M \lb r_s+r_d\rb/\lb r_sr_d\rb}/2$. When $j$ reaches its maximal value in table \ref{tab:1}, the difference in time delays for two spin directions can reach $\sim 0.22$ [s].

\begin{figure}[htp!]
\centering
\includegraphics[width=0.6\textwidth]{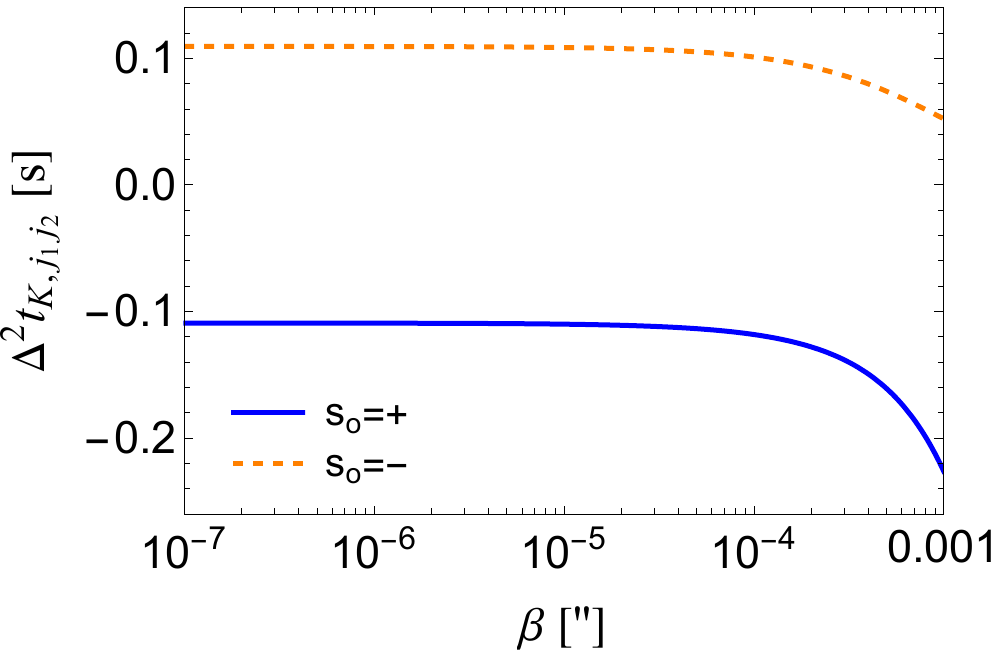}\\
(a)\\
\centering
\includegraphics[width=0.6\textwidth]{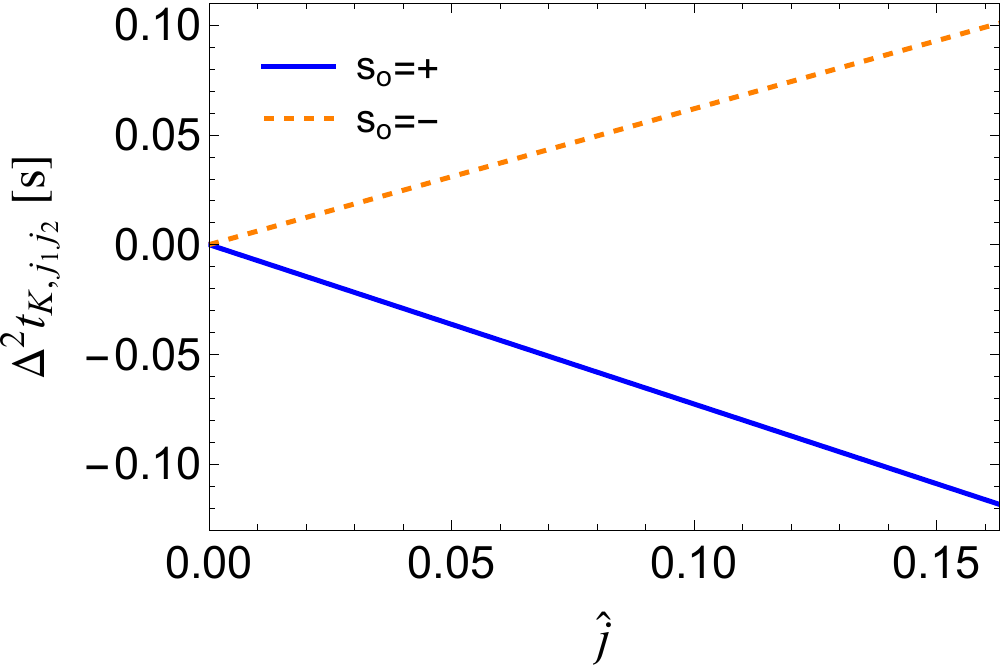}\\
(b)
\caption{The time delay \eqref{eq:kdtsj} as functions of $\beta$ (a) and particle spin $\hat{j}$ (b). We used the Sgr A* SMBH as the lens and assumed $r_s=r_{S_{39}}=1.113 \times 10 ^{-3}~\mathrm{[pc]}$, $r_d=8.1~\mathrm{[kpc]}$, $\hat{a}=0.1$ \cite{Fragione:2020khu}. Other parameters used are: $\hat{j}_1=\hat{j}_2=1.63 \times 10^{-1}$, $s_{j1}=-s_{j2}=+$ in (a) and $\beta=10^{-4}~[^{\prime\prime}]$ in (b). In (b), we varied $\hat{j}$ to $1.63 \times 10^{-1}$ according to the fourth row of table \ref{tab:1}. In both plots,
$E_\nu=10~[\mathrm{MeV}]$ and the neutrino mass 3 in table \ref{tab:1} are used to deduce the velocity of the signal.}
\label{fig:td2kerrplot}
\end{figure}

In figure \ref{fig:td2kerrplot}, we plot the time delay $\Delta^2 t_{K,j_1j_2}$  according to eq. \eqref{eq:kdtsj} for two signals with spin of same size $ j$ but antiparallel to each other $s_{j_1}=-s_{j_2}=+$. Parameters of the lens are same as in figure \ref{fig:tdkerrplot}.
From \ref{fig:td2kerrplot} (a), it is seen that as $\beta$ increases to about $10^{-4}~[^{\prime\prime}]$, time delays between different spin directions along either side of the lens are almost constant. Indeed, in this range of $\beta$, $\eta_K\gg 1$ so that
eq. \eqref{eq:kdtsj}
approaches
\begin{align}
\Delta^2 t_{K,j_1j_2} \lb v\to1, \eta_K\gg 1\rb \approx-\frac{5s_o j }{2}\sqrt{\frac{M(r_s+r_d)}{r_sr_d}}+\mathcal{O}\lb r_i\epsilon^4,~j\epsilon^2\rb. \label{eq:ktd2lim}
\end{align}
This suggests that in this case, the time delay is not only insensitive to $\beta$ but also the spacetime spin $\hat a$. On the other hand, as $\beta$ further increases beyond $10^{-4}~[^{\prime\prime}]$, then the limit \eqref{eq:ktd2lim} is broken, and
the time delay between the two signals with opposite spins but both moving in clockwise (or anti-clockwise) direction decreases (or increases) in size. This is intuitively understandable because as $\beta$ or equivalently $\varphi_0$ increases, both the two trajectories in the clockwise (or anti-clockwise) directions, regardless the particle spin, become shorter (or longer) and consequently the time delay becomes smaller (or larger). If $\beta$ continues growing so that $\eta_K\ll 1$,  then from in eq.  \eqref{eq:kdtsj}, one would have
\begin{align}
\Delta^2 t_{K,j_1j_2} \lb v\to1, \eta_K\ll 1\rb \approx\frac{3\lb 1+s_o\rb j \varphi_0}{4} .\label{eq:ktd2lim2}
\end{align}
That is, the time delay between signals with different spin directions from the counter-clockwise direction will be much larger than that from the clockwise direction. From the definition \eqref{eq:kbkdefs3} of $\eta_K$, we see that $\eta_K\ll 1$ means that $\varphi_0$ should satisfy $\varphi_0\gg 4\sqrt{M(r_s+r_d)/(r_sr_d)}$ for relativistic signals. Consequently the time delay \eqref{eq:ktd2lim2} still has a lower limit for $s_o=+$
\begin{align}
\Delta^2 t_{K,j_1j_2} \lb v\to1, \eta_K\ll 1\rb \gg6j \sqrt{\frac{M\lb r_s+r_d\rb}{r_sr_d}}.
\label{eq:dtlb}
\end{align}
That is, the solid line in figure \ref{fig:td2kerrplot} (a) at large $\beta$ will be larger than the value on the right-hand side of eq. \eqref{eq:dtlb}.

In figure \ref{fig:td2kerrplot} (b), the dependence of $\Delta^2 t_{K,j_1j_2}$ on $\hat j$ is plotted for $\beta=10^{-4}~[^{\prime\prime}]$. As suggested by \eqref{eq:kdtsj}, this time delay must be exactly zero when $\hat j$ is zero. As $\hat j$ increases, then the slope of the curve is actually also determined by \eqref{eq:ktd2lim} because
$\eta_K\gg 1$.
As $\hat j$ reaches its maximal value, then the time delay reaches about $+ 0.10$ [s]  for $s_o=-$ and $- 0.12$ [s]  for $s_o=+$ respectively.

Besides the time delay between two particles with the same $j$, we can also study the time delay between two signals with different spin sizes. For example, if one ultra-relativistic signal has spin $j$ while the other signal (not necessarily the same kind) has spin zero and they are emitted simultaneously from the same location, then the time delay between them will be half the size as in figure \ref{fig:td2kerrplot}.

\subsection{Teo wormhole spacetime results \label{subsec:teo}}

Recently, the deflection angle of spinless signals in Teo spacetime was considered in the weak field limit using the Gauss-Bonnet theorem method \cite{Jusufi:2017mav,Jusufi:2018kry,Crisnejo:2019ril} (see \cite{Li:2019qyb} for a brief review). However, the effects of signal spin on the deflection angle, apparent angles of the images as well as the time delays in this spacetime, have not been considered yet. Therefore in this section, we will apply the perturbative method developed above to this case and briefly analyze the relevant results.

The Teo wormhole metric, describing a rotating traversable wormhole \cite{Teo:1998dp}, is given by
\be
\dd s^2=-N^2\dd t^2+\Phi \dd r^2+r^2K^2\lsb \dd\theta^2+\sin^2\theta \lb\dd\varphi-\omega\dd t\rb^2\rsb ,\nn
\ee
where the metric functions are \cite{Abdujabbarov:2016efm}
\be
N=K=1+\frac{\lb 4J_Td\cos\theta\rb^2}{r},~\Phi=\lb 1-\frac{b_T}{r}\rb^{-1},~\omega=\frac{2J_T}{r^3}. \nn
\ee
Here $J_T$ is the total angular momentum of the spacetime. $b_T$ is the throat size and it can be shown to equal to two times the ADM mass $M_T$ of the spacetime \cite{Shaikh:2018kfv}, i.e., $b_T=2M_T$. Parameter $d$ is some constant to tune the dimension of the involved term and it becomes irrelevant on the equatorial plane.
Then setting $\theta=\pi/2$ on the equatorial plane, we can read off the metric functions in the form of eq. \eqref{eq:SASlinement} as
\begin{align}
A\left(r\right)=1-\frac{4J_T^2}{r^4},\quad
B\left(r\right)=-\frac{4J_T}{r},\quad
C\left(r\right)=r^2,\quad
D\left(r\right)=\lb 1-\frac{2M_T}{r}\rb^{-1}.\label{Teo}
\end{align}
Their asymptotic expansion coefficients are easily found to be
\begin{subequations}
\label{eq:Teo}
\begin{align}
&a_0=1,~a_4=-4a^2M_T^2,~a_{1,2,3,n\geq5}=0,\\
&b_0=-4aM_T,~b_{n\geq1}=0,\\
&c_0=1,~c_{n\geq1}=0,\\
&d_0=1,~d_1=2M_T,~d_2=4M_T^2,~\cdots,
\end{align}
\end{subequations}
where we have set $a=J_T/M_T$ as the angular momentum per unit mass of the wormhole.

Substituting eq. \eqref{eq:Teo} into \eqref{eq:ynexp}, the first few $y_n$ coefficients of the deflection angle in Teo wormhole spacetime are obtained as
\begin{subequations}
\label{eq:Teoexamp}
\begin{align}
y_{T,0}=&s_o,\\
y_{T,1}=&M_T\lb s_o -\frac{2a}{b v}\rb-\frac{s_jj}{2 b v},\\
y_{T,2}=&M_T^2\lsb\frac{3s_o  }{2}-\frac{4a}{b v}+\frac{12s_o a^2 }{b^2 v^2}-\frac{3s_jj}{b v}\lb 1-\frac{4s_o  a }{b v}\rb-\frac{3s_o j^2}{b^2 v^2}\rsb.
\end{align}
\end{subequations}
Substituting into eq. \eqref{eq:dphifinal}, the deflection angle in this spacetime becomes
\be
\Delta\varphi_T=\sum_{i=s,d}\sum_{n=0}^\infty y_{T,n}\frac{I_n \lb \theta_i \rb}{b^n}.
\ee
In the large $r_{s,d}$ limit, this result can be expanded into a dual power series of $\lb M/b\rb$ and $\lb b/r_{s,d}\rb$. Or more easily, substituting directly the coefficients \eqref{eq:Teo} into eq. \eqref{aformula} yields this expansion
\begin{align}
\label{eq:dft}
\Delta\varphi _{T}=\sum_{i=s,d}s_o\left\{\frac{\pi  }{2}+\frac{M_T}{b}-\frac{b }{r_i}+\lsb\frac{3 \pi  }{8}-\frac{s_o \lb 2 \hat{a}+s_j\hat{j} \rb}{ v}\rsb\lb\frac{M_T}{b} \rb^2+\mathcal{O} \left(\epsilon^3\right) \rcb,
\end{align}
where as before, we have introduced $\hat{j}=j/M_T$ and $\hat{a}=a/M_T$. Setting $\hat j=0$, this
reduces to eq. (68) of ref. \cite{Li:2019qyb} for spinless signals in this spacetime.
Note that to the leading order $\mathcal{O}\lb M_T/b\rb$, deflection \eqref{eq:dft} is only half the size of the classical Schwarzschild spacetime result $4M_T/b$. This is because the temporal metric component does not have nonzero $\mathcal{O}\lb 1/r\rb$ order expansion coefficient. Similarly, difference in higher order terms compared to the Kerr spacetime deflection \eqref{eq:dfk} are also caused by the difference between asymptotic coefficients in \eqref{eq:Teo} and \eqref{eq:Kerr}.

Using \eqref{expandangle}, we can also obtain the apparent angles in the Teo wormhole spacetime as
\be
\theta_{T,s_o s_j}=\frac{b_{T,0s_o}}{r_d}
+\frac{b_{T,1s_os_j}}{r_d}+\mathcal{O}\lb \epsilon^3\rb,
\label{eq:thetateo}
\ee
where
\begin{subequations}
\begin{align}
b_{T,0s_o}=&\frac{\varphi_0 r_d r_s}{2\lb r_d+r_s\rb}\lb \sqrt{ 1+\eta_T} -s_o \rb,\\
 b_{T, 1 s_o s_j}=&\frac{\eta_T \lsb -8 s_o\lb 2a+s_j j \rb +3M \pi v \rsb}{32v \sqrt{1+\eta_T }\lb  \sqrt{1+\eta_T}-s_o \rb},\label{eq:tb1}\\
\eta_T=&\frac{8M\lb r_d +r_s \rb}{\varphi^2_0 r_d r_s }.
\end{align}
\end{subequations}
Again, because of the difference in the asymptotic coefficients in \eqref{eq:Teo} and \eqref{eq:Kerr}, the impact parameters above are also slightly different from their counterpart eq. \eqref{eq:kbkdefs} in the Kerr spacetime case.

Now the total travel time in the Teo wormhole spacetime is still given by eq. \eqref{eq:dtfinal}
\begin{align}
\Delta t_T =\sum_{i=s,d}\sum_{n=-1}^\infty z_{T,n}\frac{I_{n-1} \lb \theta_i \rb}{b^n},
\label{eq:ttteo}
\end{align}
but with new coefficients $z_{T,n}$ obtained by substituting the asymptotic coefficients \eqref{eq:Teo} into eq. \eqref{eq:timedelay}
\begin{subequations}
\begin{align}
\label{eq:extteo}
z_{T,-1}=&\frac{1}{v},\\
z_{T,0}=&\frac{M_T}{v},\\
z_{T,1}=&\frac{M_T^2}{v}\lsb\frac{3}{2}- \frac{2s_o a }{b v} -\frac{s_jj}{b v} \lb\frac{3s_o   }{2}-\frac{4a }{b v}\rb + \frac{ 4a^2}{b^2 v^2}+\frac{j^2}{b^2 v^2}\rsb,\\
z_{T,2}=&\frac{M_T^3}{v}\lsb\frac{5}{2}-\frac{2a}{bv}\lb 3s_o-\frac{6 a}{bv}+\frac{16 s_o a^2 }{b^2 v^2} \rb -\frac{s_j j}{b v}\bigg( 5 s_o-\frac{18 a }{b v}+\frac{48 s_o a^2 }{b^2 v^2} \bigg)\rsb \nn\\&+\frac{6j^2}{b^2v^2}\lb 1- \frac{4s_o a }{b v}\rb-\frac{4 s_o s_j j^3}{b^3 v^3}- M_T\lsb2s_o a b-\frac{s_j aj}{v}\lb 3+\frac{1}{v^2}\rb+\frac{j ^2}{v} \lb1-\frac{8 s_o a }{b v}\rb\rsb.
\end{align}
\end{subequations}
In the large $r_{s,d}$ limit, to the leading order eq. \eqref{eq:ttteo} becomes 
\begin{align}
\label{tteo}
\Delta t_T=&\sum_{i=s,d} \lcb\frac{r_i}{v}-\frac{1}{2 v} \frac{b^2}{r_i}+\lsb\frac{3 \pi }{4v}-2s_o \hat{a}\lb 1+\frac{1}{v^2} \rb-\frac{s_o s_j \hat{j}}{v^2}\rsb \frac{M_T^2}{b}\right.\nn\\&+\frac{M_T}{v} \ln \left(\frac{2 r_i}{b}\right)+\mathcal{O}\lb r_i\epsilon^4,~j\epsilon^2\rb\Bigg\}.
\end{align}

Finally, the time delay $\Delta t^2_{T,o}$ between signals from different sides of the lens and $\Delta t^2_{T,j_1j_2}$ between different spins are given by
\begin{align}
  \Delta^2 t_{T,o} = &\ln \lb \frac{\sqrt{\eta_T+1}+1 }{\sqrt{\eta_T+1}-1 }\rb \cdot \frac{2 M }{v}  +\frac{\sqrt{\eta_T+1}}{\eta_T}\cdot \frac{4 M}{ v}-2a\varphi_0\sqrt{\eta_T+1}-\frac{3\pi M \varphi_0}{32 v}\nn\\&-\lb2a+s_j j\rb\varphi_0 \lb\frac{5\eta_T+6}{8v^2\sqrt{\eta_T+1}}\rb+\mathcal{O}\lb r_i\epsilon^4,~j\epsilon^2\rb,\label{timedelayinTeo}  \\
  \label{timedelay2Teo}
\Delta^2 t_{T,j_1j_2}=&\frac{\lb s_{j2}j_2-s_{j1}j_1\rb\varphi_0}{16 v^2}\lsb6+\frac{s_o\lb5\eta_T+6\rb}{\sqrt{\eta_T+1}}\rsb+\mathcal{O}\lb r_i\epsilon^4,~j\epsilon^2\rb.
\end{align}
We see that these time delays have a similar structure as their counterparts, eqs. \eqref{timedelayinkerr} and \eqref{eq:kdtsj}, in Kerr spacetime, but with simpler coefficients.

\section{Conclusions \label{sec:conc}}

In this work, we developed a perturbative method to solve the deflection angle, lensing equation and time delay for spinning particles moving in the equatorial plane of arbitrary stationary and axisymmetric spacetimes. This method works in the weak field limit and takes into account the finite distance effect of the source and detector. The result for the deflection angle $\Delta\varphi$, as given in eq. \eqref{eq:dphifinal}, takes a quasi-power series form of $\lb M/b\rb$. It can also be expanded into a dual power series of $\lb M/b\rb$ and $\lb b/r_{s,d}\rb$. The particle spin $j$ affects the deflection from order $\lb M/b\rb^2$, as seem from eq. \eqref{aformula}, or from eqs. \eqref{eq:dfk} and \eqref{eq:dphihexp} in the Kerr spacetime case. A similar series form for the total travel time is also obtained.
These results are used to find the apparent angles \eqref{expandangle} and time delays \eqref{timedelaycase1} and \eqref{timedelay2} of the GL images in a perturbative way.

As a perturbative work, it is  essential to find out the orders at which each kind of coupling involving particle spin $j$ appears in the main quantity studied, i.e., the deflection $\Delta\varphi$. Therefore let us make a detour to comment in this regard. Previous studies have mentioned at least the following kinds of couplings involving $j$: the particle's spin-orbital coupling, the particle's spin and total angular momentum coupling, the particle's spin-spin coupling, the particle-spacetime spin-spin coupling. The former three are between different kinds of momenta of the test particle and the last is between the signal and the spacetime. First of all, we found that if $\Delta\varphi$ is computed to high enough orders, all these couplings will appear, as seen from eq. \eqref{eq:dphihexp}. Using definition \eqref{eq:oribtangdef} for the orbital angular momentum, all terms in eq. \eqref{eq:dphihexp} that are proportional to $(s_os_j \hat{j})^1$ can be transformed to $ s_j\sqrt{1-v^2}l\hat{j}/(mvb)$, which is the spin-orbit coupling. Because of this extra $1/b$ introduced in this transform, the $l\hat{j}$ coupling terms will only appear in order $\lb M/b\rb^3$ or above in $\Delta\varphi$. Similarly, one can also use the relation \eqref{eq:Lrinfinity} between $L,~l$ and $j$ to transform the same $s_os_j j$ terms to be proportional to $\lb \sqrt{1-v^2}L-s_jjm\rb\hat{j}/b$. Thus we conclude that the coupling between spin and total angular momenta of the particle and the particle's spin-spin coupling appear from the same $\lb M/b \rb ^3$ order, as the spin-orbital coupling does. Inspecting the $\lb M/b\rb^3$ order in eq. \eqref{eq:dphihexp}, we also see that the $s_j\hat{j}\hat{a}$ term appears, which is the particle spin-spacetime spin coupling term. Therefore, secondly, we found all these couplings involving $\hat{j}$, when written in a product form, actually happen at the $\lb M/b\rb^3$ order or above. 

The series formula for the deflection angle, total time, apparent angles and time delays are then applied to the Kerr spacetime and the results are analyzed carefully. It is found that in general, the series result for the deflection angle converges very well to the true physical value as the series order increases. We took neutrinos with small mass as example of spinning signals because of their large spin to mass ratio $j$. It is found that if the spin angular momentum of the signal is parallel (or antiparallel) to its orbital angular momentum, then the size of the deflection angle is decreased (or increased), as illustrated in figure \ref{fig:jtodphi}.
The apparent angles of the images in GL are roughly affected by $j$ in the same way as the deflection angle. However, since the contribution of $j$ to the apparent angles is suppressed by its ratio to the lens mass, only for lighter lens and very large $j$ their contribution might reach the $\sim 1~[^{\prime\prime}]$ order level.

The time delay in Kerr spacetime between signals with different spins is found in eq. \eqref{eq:kdtsj}. This is proportional to the spin difference between the signals from the leading order. Therefore measuring this time delay could be used to constrain the signal's spin to mass ratio $j$. For Sgr A* SMBH, depending on the value of $j$ and source's location coordinates, this time delay could reach the $\mathcal{O}\lb 0.1\rb~\mathrm{[s]}$ level. For neutrinos, this implies that if $j$ is large enough, then measuring this time delay could shed light on their masses.

We also applied the perturbative method, including the deflection angle, apparent angles and time delays, to the Teo wormhole spacetime without any difficulty. Actually we have shown that the method and results developed here can be used to the equatorial motion of spinning particles in arbitrary stationary and axisymmetric spacetimes whose metric functions allow power series asymptotic expansions.
We emphasize that these include arbitrary static and spherically symmetric spacetimes too and therefore our result can also be directly applied to those by simply setting
the coefficients $b_n=0$ in section \ref{sec:dphianddt} and \ref{sec:glgeneral}.

For extension of the work, one important point to consider is to allow the spin direction to be non-perpendicular to the plane of the motion. Although it is known that the signal with its spin perpendicular to the equatorial plane can remain in the plane and keep its spin direction, this is certainly not the most general scenarios of particle's spin orientation (nor its motion). Considering all the spin directions should help to reveal new features or effects of spin on the deflection angle or GL of spinning particles.

\acknowledgments

We thank Tingyuan Jiang and Haotian Liu for their helpful discussions. This work is supported by the MOST China 2021YFA0718500.

\appendix

\section{Integration formulas and their expansions \label{sec:app1}}

Integration in eq. \eqref{I} can be carried out using a change of variables $u=\sin\xi$ and their results are found in ref. \cite{Xu:2021rld}. Here we quote them directly
\begin{align}
I_n\lb\theta_i\rb \equiv&  \int_{\theta_i}^{\pi/2} \sin^{n} \xi \dd \xi  \\
 =& \begin{cases}
\displaystyle  \cot\theta_i,&n=-2,\\
\displaystyle  \ln\lsb\cot\lb\frac{\theta_i}{2}\rb\rsb,&n=-1,\\
\displaystyle  \frac{\lb n-1\rb!!}{n!!} \left(\frac{\pi}{2}-\theta_i
+\cos\theta_i\sum_{j=1}^{[n/2]} \frac{\lb2j-2\rb!!} {\lb2j-1\rb!!}\sin^{2j-1} \theta_i\right),&n=0,2,\cdots,\\
\displaystyle  \frac{\lb n-1\rb!!}{n!!} \cos\theta_i \left(1
+\sum_{j=1}^{[n/2]} \frac{\lb2j-1\rb!!}{\lb 2j\rb!!} \sin^{2j}\theta_i\right),&n=1,3,\cdots.
\end{cases}
\label{eq:inres}
\end{align}
In the $\theta_i=0$ limit, i.e., the infinite source/detector distance limit, $I_n$ for non-negative $n$ can be further simplified to 
\be
I_n\lb0\rb=
\frac{\lb n-1\rb!!}{n!!}\begin{cases}
\displaystyle  \frac{\pi}{2},&~n=0,~2,~\cdots,\\
\displaystyle  1,&~n=1,~3,~\cdots.
\end{cases} \label{eq:inreslim}
\ee

Since in the weak field limit $r_{s,d}\gg b\gg M$, using eq. \eqref{eq:thetares} and asymptotic expansions \eqref{eq:ABCD},
$\theta_i$ can be written as dual series of $\lb M/b\rb$ and $\lb b/r_i\rb$
\begin{align}
\theta_i =&\frac{b}{r_i}+\frac{b}{2r_i^2}\lb \frac{a_1}{v^2}-c_1\rb
+\frac{b^3}{6r_i^3}
-\frac{s_o}{2r_i^2}\lsb \frac{ b_0}{v}-\frac{s_j j \lb a_1+d_1\rb}{v}\rsb
+\mathcal{O} \lb \epsilon^5 \rb .
\label{seriestheta}
\end{align}
In a typical GL by SMBH, roughly $M/b$ and $b/r_i$ are same order infinitesimals and in this case the second and third terms above are of order three while the fourth term is of order four.
Substituting eq. \eqref{seriestheta} into  \eqref{eq:inres} and further expanding in small $\lb M/b\rb$ and $\lb b/r_i\rb$, the first few $I_n$ can be shown to also take the dual power series form
\begin{subequations}
\label{eq:saslnexp}
\begin{align}
I_{-2}=&\sum_{i=s,d}\lcb \frac{r_i}{b}-\frac{b}{2 r_i} + \frac{1}{2b}\lb  c_1-\frac{a_1}{v^2}\rb+\frac{s_o}{2b^2v} \lsb b_0- s_j j \lb a_1 +d_1 \rb\rsb+ \mathcal{O} \lb\epsilon^3\rb \rcb, \\
I_{-1}=&\sum_{i=s,d} \lcb \ln \lb \frac{2r_i}{b} \rb +\frac{1}{2r_i}\lb c_1-\frac{a_1}{v^2 }\rb+\frac{s_o}{2br_iv}\lsb b_0 - s_j j\lb a_1+d_1 \rb\rsb+ \mathcal{O}\lb\epsilon^2\rb\rcb, \\
I_0=&\sum_{i=s,d} \lcb\frac{\pi}{2}-  \frac{b}{r_i} + \mathcal{O}\lb\epsilon^3\rb\rcb, \\
I_1=& \sum_{i=s,d}\lcb 1 + \mathcal{O}\lb\epsilon^2\rb \rcb,\\
I_2=& \sum_{i=s,d} \lcb \frac{\pi}{4} + \mathcal{O}\lb\epsilon^3\rb \rcb,\\
I_3=& \sum_{i=s,d}\lcb \frac{2}{3} + \mathcal{O}\lb\epsilon^4\rb \rcb.
\end{align}
\end{subequations}

\section{Higher order terms in $\Delta\varphi_K$ \label{sec:highphik}}

Here for potential future reference, we present the deflection angle in Kerr spacetime, $\Delta \varphi_K$, to orders higher than eq. \eqref{eq:dfk}
\begin{align}
\Delta\varphi_K=&\sum_{i=s,d} s_o \lcb \frac{ \pi  }{2}-\frac{b }{r_i}+\lb 1+\frac{1}{v^2}\rb\lb\frac{M}{b}\rb+\left[\frac{3\pi}{2}\lb\frac{1}{4}+\frac{1}{v^2}\rb-\frac{2 s_o \hat{a} }{ v}-\frac{2s_os_j \hat{j}}{ v}\right]\lb\frac{M}{b}\rb^2\right.\nn \\
&-\frac{1}{2}\lb1-\frac{1}{ v^2}\rb\lb \frac{M}{b}\rb \lb \frac{b}{r_i}\rb ^2 +\lcb\frac{5 }{3 }+\frac{15 }{ v^2}+\frac{5 }{ v^4}-\frac{1}{3 v^6}-\frac{s_o\pi \hat{a}}{v}\lb3+\frac{2  }{v^2}\rb\right.\nn\\&\left.+ \hat{a}^2\lb 1+\frac{ 1}{ v^2}\rb-s_j\hat{j} \lsb\frac{3s_o\pi }{2v}\lb \frac{3}{2}+\frac{1}{  v^2}\rb-2\hat{a}\lb 1+\frac{1}{ v^2}\rb\rsb\rcb\lb\frac{M}{b}\rb^3-\frac{ 1 }{6 }  \lb \frac{b}{r_i}\rb^3\nn\\&+\lcb\frac{105\pi}{8}   \left(\frac{1}{16}+\frac{1}{v^2}+\frac{1}{ v^4}\right)-\frac{6s_o\hat{a}}{v} \left(5+\frac{10 }{ v^2}+\frac{1}{ v^4}\right)+\frac{3\pi\hat{a}^2}{4}\lb \frac{15}{8}+\frac{9 }{v^2}+\frac{1}{  v^4}\rb\right.\nn\\&-\frac{ 2s_o\hat{a}^3 }{ v}-s_j \hat{j}\lsb\frac{4s_o}{v}\lb5+\frac{10}{ v^2}+\frac{1}{ v^4}\rb-\frac{3\pi\hat{a}}{2}\lb\frac{3}{2}+\frac{15}{2 v^2}+\frac{1}{  v^4}\rb+\frac{6s_o \hat{a}^2 }{ v}\rsb \nn\\
&\left.\left.-3\pi\hat{j}^2\lb 1+\frac{4}{ v^2}\rb\rcb\lb\frac{M}{b}\rb^4-\frac{s_o\hat{a}}{v}\lb \frac{M}{b}\rb^2  \lb \frac{b}{r_i}\rb^2+\mathcal{O} \left[\lb\frac{M}{b}\rb^5,\lb\frac{b}{r_{i}}\rb^5\right]\rcb.\label{eq:dphihexp}
\end{align}
Note that terms involving the spacetime and signal spin coupling  $\hat{j}\hat{a}$ start to appear from the third order, i.e., the $\mathcal{O}\lb M/b \rb^3$ order.


\end{document}